\documentclass[a4paper,11pt]{amsart}

\usepackage{amsmath,graphicx,amsfonts,amssymb,amsthm,url,bbm,courier,empheq,newtx} 

\usepackage{colortbl,booktabs,amsaddr}
\usepackage{xcolor}
\usepackage[utf8]{inputenc}

\usepackage{natbib}

\newtheorem{theorem}{Theorem}[section]%  
\newtheorem{proposition}[theorem]{Proposition}%
\newtheorem{corollary}[theorem]{Corollary}

\newtheorem{lemma}[theorem]{Lemma}% 

\theoremstyle{remark}
\newtheorem{remark}{Remark}%

\theoremstyle{definition}

\newcommand{\btheta}{\boldsymbol{\theta}}
\newcommand{\gr}{\mathrm{graph}}

\DeclareMathOperator{\trace}{trace}

\begin{document}
\title[Estimation under the Emax Model]{Maximum Likelihood Estimation under the Emax Model: Existence, Geometry and Efficiency}

\author{Giacomo Aletti}
\address{ADAMSS Center, Università degli Studi di Milano, 
Milan 20133, MI, Italy}
\email{giacomo.aletti@unimi.it}

\author{Nancy Flournoy}
\address{Department of Statistics, University of Missouri, Columbia 65211, MO, United States of America}
\email{flournoyn@missouri.edu}

\author{Caterina May}
\address{Dipartimento di Studi per l'Economia e l'Impresa, Università del Piemonte Orientale, Novara 28100, NO, Italy}
\address{Department of Mathematics, King's College London, London WC2R 2LS, Strand, United Kingdom}
\email{caterina.may@uniupo.it}

\author{Chiara Tommasi}
\address{Department of Economics, Management and Quantitative Methods, Università degli Studi di Milano, 
Milan 20122, MI, Italy}
\email{chiara.tommasi@unimi.it}


\begin{abstract}
This study focuses on the estimation of the Emax dose-response model, a widely utilized framework in clinical trials, agriculture, and environmental experiments. Existing challenges in obtaining maximum likelihood estimates (MLE) for model parameters are often ascribed to computational issues but, in reality, stem from the absence of a MLE. Our contribution provides a new understanding and control of all the experimental situations that practitioners might face, guiding them in the estimation process.
We derive the exact MLE for a three-point experimental design and we identify the two scenarios where the MLE fails. To address these challenges, we propose utilizing Firth's modified score, providing its analytical expression as a function of the experimental design. Through a simulation study, we demonstrate that, in one of the problematic cases, the Firth modification yields a finite estimate. For the remaining case, we introduce a design-augmentation strategy akin to a hypothesis test.
\end{abstract} 




\maketitle

\tableofcontents

\section{Introduction}%\label{sec:intro}
The Emax model is   well-characterized in the literature and  widely used  in a variety of fields including clinical trials, biochemistry, agriculture and environmental experiments 
\citep[e.g.,][]{bretz2010practical, baker2010full,baker2016mathematical,han2011comparison, holford1981understanding,macdougall2006,denney2017pharma,rath2022application}.
 The Emax model is an extension of the widely used Michaelis-Menten model \citep{michaelis1913kinetik}. 
In the literature, it is commonly presented with three or four parameters (referred to as the four-parameter logistic model). In this study, we focus on the three-parameter version, which is frequently used in dose-response studies, where it is often assumed that the response mean can be described by a simple concave function that increases monotonically with a covariate $x\in\mathcal{X}$, such as dose or stress 
[see \cite{LeonMill2009} and \cite{Leonov23}].  
Other aspects of this model have been studied by \citet{Dette:2010}, \citet{dragalin2010simulation}, and \citet{Flou:May:Toma2021}. 

Although the maximum likelihood estimates (MLEs) of  the model parameters are asymptotically consistent, it is well-known that uncertainty in finite samples may lead to the non-existence of the MLE. There also is a large literature on other  convergence problems associated with Emax parameters estimation algorithms; see, for example, \citet{fedorov2013optimal}, \citet{flournoy2020performance} and \citet{Leonov23}.
We analyze the same convergence problems theoretically (not from a numerical point of view), exposing  geometric features important to the application of  likelihood methods under the Emax model. 

Our main concern is to guarantee (as much as possible) a finite estimate of the model parameters. It is well known that the D-optimal design (which has three support points) leads to the most precise parameter estimate. Herein, we show that (among all the three-point designs) it almost minimizes the probability of the MLE non-existence, that is a very useful property.
For analytical and geometrical tractability, we  focus only on three-point experimental designs. Three-point designs are  common choice in experimental practice; in addition, they lay the foundation for extensions to more complex designs. Provided that observed means  at the design's support points have an increasing concave shape, we are able to give an analytic expression of the MLE vector as a function of the data. We  prove that if the observations do not satisfy this geometric shape, then the MLE  does not exist. In particular, 
we identify two different scenarios (with several sub-cases) where the MLE fails. We call these problematic scenarios as \lq\lq Case 1'' and \lq\lq Case 2'' and provide the probabilities of observing them.  
In these unlucky cases, to solve the issue of non-existent MLEs,  we recommend to estimate the parameters with the roots of the modified score equations \citep[see][]{firth1993bias}. 
Briefly,
\citeauthor{firth1993bias} developed a general method to reduce  the  bias of the MLE by modifying the score function. The roots of Firth's modified score equations result in first-order unbiased estimators, herein called \textit{Firth-modified estimators}, \citep[see also][]{kosmidis2009bias, kosmidis2009biasb}. But, more importantly for Emax model estimation,  \citet[section 3.3]{firth1993bias} shows that his  estimators may circumvent 
the problem of the non-existence  MLEs even with moderate sample sizes.

To apply Firth's method, we derive the analytical expression of  Firth's correction for the score function of the Emax model as a function of any given design. We find that  Firth's modified score leads to a finite estimate only in Case 2.  
Unfortunately, in Case 1 Firth's method fails. 
In this later scenario, however, a geometric argumentation leads  to  a proposal of design augmentation, that consists in identifying the region where an additional experimental point should be added in order to obtain a finite estimate. Let us note that this design-correction is proved to be equivalent to a hypothesis test on the most critical  parameter of the Emax model. 

In Section~\ref{sec:notation}, we give the notation and common parameterization of the Emax model. 
Section~\ref{sect:MLE} provides the analytic expression of the MLE with  conditions for its existence and descriptions of  the bad scenarios for which the MLE does not exist.
In Section~\ref{sect:Firth}, we provide the explicit formula for the Firth correction of the Emax score function.
In Section~\ref{sec:simulation} we face the bad scenarios, showing  when Firth method succeeds as well as when a design strategy is  necessary to control the problem. {Finally Section~\ref{sect:concl}, which concludes the paper, provides  practical guidelines for an experimenter resulting from our theoretical results. }

%%%%%%%%%%%%%%%%%%%%%%%%%% 
\section{Model and Notation}
\label{sec:notation}
 The E{max} model  is $y=\eta(x,\btheta)+\varepsilon$, where $y$ denotes a response at the dose $x\in \mathcal{X}=[a,b]$;  $a\geq 0$ and $b\geq a$ are the lowest and the highest admissible doses; $\varepsilon$ is a Gaussian random error; and  $\btheta=(\theta_0, \theta_1,\theta_2)^T$ is a vector of unknown parameters that belongs to {a} {parameter space} that makes the response mean 
 \begin{equation}\label{eq:E_max}
\eta(x,\btheta)=\theta_0+\theta_1\,\frac{x}{x+\theta_2}
\end{equation}
{an increasing and concave curve} (see Section \ref{sect:param}). 

  An  experimental design is  a finite discrete probability distribution over $\mathcal{X}$:
\begin{equation}\label{xi}
\xi=\left\{\begin{array}{ccc}
x_1 &\cdots & x_M \\
\omega_1 & \cdots & \omega_M
\end{array}
 \right\},
\end{equation}
where $x_i$ denotes the {$i$th experimental point, or treatment dose, that may be} used in the {study} and $\omega_i$ {is} the proportion of {experimental} units to be taken at  {that} point{; $\omega_i\geq 0$ with  $\sum_{i=1}^{M}\omega_i=1$,} $ i=1,\ldots,M$ and $M$ is finite.
 
 Assuming that at the dose $x_i$ we observe $n_i=\omega_i \cdot \,n$ independent responses, $y_{i1},\ldots,y_{in_i}$  (for $i=1,\ldots,M$), the most common estimate of $\btheta$ is the maximum likelihood estimator (MLE).
 It is well known that, to improve the precision of the MLE, we should apply a D-optimal design, since it minimizes the generalized asymptotic variance of the MLE for $\btheta$. General references on optimal design of experiments include \citet{Fedo:Theo:1972}, \citet{Atk2007} and \citet{silvey2013optimal}, while \citet{fedorov2013optimal} is specific to optimal designs for nonlinear response models. In particular, \citeauthor{fedorov2013optimal} describe strategies for implementing D-optimal designs when they are functions of the unknown parameters.  We recommend the use of sequential adaptations for the Emax model in Section~\ref{sect:concl}. 

\noindent Let
  \begin{equation*}
%\label{I}
 I(\xi;\btheta)=\int_\mathcal{X} \nabla \eta(x,\btheta)
 \nabla \eta(x,\btheta)^T d\xi(x)
 \end{equation*}
denote the Fisher information matrix of an experiment with design $\xi$ under model \eqref{eq:E_max}, where
 $\nabla \eta(x,\btheta)$ denotes the gradient of the mean response $\eta(x,\btheta)$ with respect to $\btheta$. The D-optimal design for the Emax model  \citep[provided by][]{Dette:2010}  is
 \begin{equation}\label{eq:D-opt}
 %\label{D-opt}
 \xi^*(\btheta)= \arg\max_{\xi\in{\Xi}}\ {\rm Det}[{I}(\xi ,;\btheta)]
=\left\{
\begin{matrix}
a & x^*(\theta_2) & b \\
1/3 & 1/3 & 1/3
\end{matrix}
\right\},
\end{equation}
where $\Xi$ is  the set of all possible designs and
\begin{equation}
\label{eq:D_opt_dose}
x^*(\theta_2)=\frac{b(a+\theta_2)+a(b+\theta_2)}{(a+\theta_2)+(b+\theta_2)}.
\end{equation}  
 Design $\xi^*({\btheta})$ is said to be locally optimal because it depends on the unknown parameter value ${\btheta}$, due to the non-linearity of $\eta(x,\btheta)$. See section~\ref{sect:concl} for implementation recommendations.

 \subsection{Parameter Space with Interpretation}\label{sect:param}
{It is important to note for what will be discussed in this section  that  %if the Emax response mean is given by \eqref{eq:E_max}, then 
the equation $y=\eta(x,\btheta)$ given by model \eqref{eq:E_max} on the Cartesian plane is an hyperbola  with the upper horizontal asymptote $y=\theta_0+\theta_1$ and the vertical asymptote $x=-\theta_2$. The Emax response mean curve is the concave branch of this hyperbola. An example (discussed further in Section~\ref{sect:MLE}) is shown in Figure~\ref{fig:initial_plot}.

\begin{figure}[htbp]
    \centering
    \includegraphics[width = .75\linewidth]{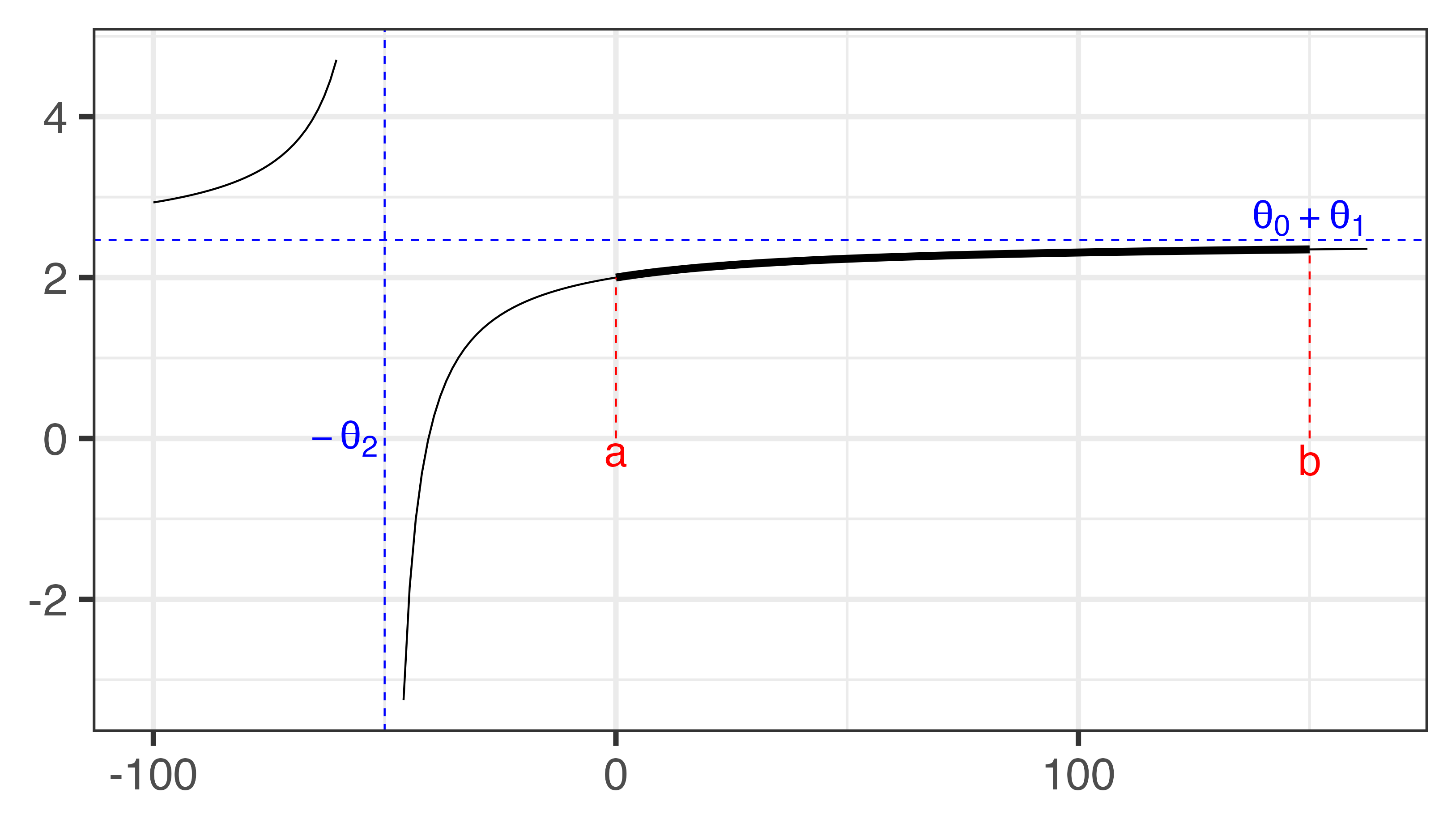}
    \caption{Emax response mean curve as part of a branch of a hyperbola. Black line: hyperbola with equation $\eta(x) = \theta_0 + \theta_1x /(x+\theta_2) = 2 + 0.467 x /( x + 50)$. Blue dashed lines: vertical ($x=-\theta_2$) and horizontal ($y= \theta_0+\theta_1$) asymptotes of the
    hyperbola. The thick part of the graph corresponds to the Emax response mean curve, which is given on the support $\mathcal{X} = [a,b]$.}
    \label{fig:initial_plot}
\end{figure}

It is {often} natural to assume %in {many} situations 
that the  dose $x=0$ is the lowest dose admissible, that is, the lower limit of the experimental space is $a=0$.
In the Emax model, if $a=0$, then   $\theta_0$ represents the response at  dose $x=0$ (placebo effect), $\theta_1> 0$ is the asymptotic maximum effect attributable to the drug (for an infinite dose) and $\theta_2>0$ is the dose which produces the half of the maximum effect. {By considering this parameter space ($\theta_1> 0$, $\theta_2> 0$) for $x \in [0,b]$, all the concave branches of the hyperbola can be fitted by the model.} 

To solve the computational problems and to ensure the existence of the MLE, several authors have bounded the space of the non-linear parameter $\theta_2$ in a {positive} compact set; for instance, \cite{Dette:2012} consider $\theta_2 \in [0.015; 1500]$. 
However, this approach excludes some parameter values that could produce a response curve which is more closely aligned with the observed data.
  In this paper we solve all the computational problems by providing an analytic solution of the MLE  for $\theta_2>0$, together with the exact conditions of its existence (see Section \ref{sect:MLE}).
  %In section \ref{sect:MLE} an exact solution of the MLE is given for $x \in[0,b]$ and for any $\theta_2>0$.

  When the lowest dose admissible is $a> 0$, the parameter interpretation given above fails. In addition, if {$a>0$ and} $\theta_2 > 0$, all  admissible response curves with vertical asymptote in $(0,a)$ are excluded. These curves may better approximate the observed data, and hence the parametric domain {of $\theta_2$ should} be extended %{also} to {$-a<\theta_2<0$ } 
  to ensure best fit. This is motivated in detail in Section~\ref{sect:geometry} by introducing a {suitable reparametrization} of the model. {The conditions of existence and the analytic solution for the MLE are provided in Section \ref{sect:MLE}, including for this case.}

\subsection{Reparametrization when the Lower Boundary of Admissible Doses is $a>0$}
\label{sect:geometry}
In some situations, {due to ethical concerns, patient exposure to placebo is not feasible}, and then the mean response curve $E(Y \vert X= x)$ is considered in a domain $\mathcal{X}=[a,b]$, $a>0$. 
Model \eqref{eq:E_max} can be rewritten as
\begin{equation}\label{eq:repEMAX}
E(Y \vert  {X}= x) = \eta(x-a,\tilde{\btheta})=\tilde{\theta}_0+\tilde{\theta}_1\,\frac{(x-a)}{(x-a)+\tilde{\theta}_2},
\end{equation}
where
\begin{equation}\label{eq:norm_tilde}
\tilde{\theta}_0 = {\theta}_0 +a \frac{{\theta}_1}{{\theta}_2 + a},
\qquad
\tilde{\theta}_1 = \theta_1  - a \frac{\theta_1}{{\theta}_2 + a},
\qquad
\tilde{\theta}_2 = \theta_2 + a.
\end{equation}
Model \eqref{eq:repEMAX} {represents an Emax model in a new parametrization, with a change in the coordinate to $\tilde{x} = x-a$} {\citep[This  same parametrization has been considered by][]{Leonov23}.}
% We can call this  \emph{natural geometry} (natural coordinates and natural reparametrization), since
It is reasonable to expect an experimenter to be interested in estimating the new parameters since $\tilde{\theta}_0$ is the mean response for the lower dose $x=a$; 
$\tilde{\theta}_1$ is the asymptotic maximum effect attributable to the drug with respect to the response at the minimal dose; and $\tilde{\theta}_2$ is the dose added to $a$ which produces the half of the maximum effect.
Likewise in the case $a=0$, the {constraints $\tilde{\theta}_1>0$, $\tilde{\theta}_2>0$} identify all the increasing and concave branches of hyperbola. 

Next Proposition~\ref{rem:cpt} shows that the {Emax} mean response curves  are not a compact set; roughly speaking, some response curves obtained as limit of curves in model \eqref{eq:E_max} do not belong to  {the same} model. As a consequence, there are situations in which the data are such that it is not possible to find an \lq\lq Emax'' mean response curve that maximizes the likelihood. %since the likelihood 
 Instead the likelihood is maximized by one of the three limiting cases listed in Proposition~\ref{rem:cpt}. {Consider now $\mathcal{X}=[a,b]$,  $a\geq 0$, and the reparametrization \eqref{eq:repEMAX} (which coincides with the standard parametrization when $a=0$).} 

\bigskip

\begin{proposition}
    \label{rem:cpt}
The set of increasing and concave branches of hyperbola on $\mathcal{X}$ is not 
locally compact in the set of bounded functions on $\mathcal{X}$ with point-wise convergence.
In fact, the limit class contains other bounded functions on $\mathcal{X}$, which precisely are
\begin{enumerate}
    \item
    \label{illcase1} the strictly increasing straight lines $\tilde{y} = m\,\tilde{x}+q$, as limits of $\eta(\tilde{x},\tilde{\btheta})$ {when}, for instance, $\tilde{\theta}_{0} = q$, $\tilde{\theta}_{1} = m\tilde{\theta}_{2}$ and
    $\tilde{\theta}_{2} \to \infty$;
    \item
    \label{illcase2} {the} horizontal lines $\tilde{y} = q$, {when} $\tilde{\theta}_{0} = q$ and \emph{a)} $\tilde{\theta}_{1}=0$ and any value of $\tilde{\theta}_{2}$ or \emph{b)} any value of $\tilde{\theta}_{1}$ and $\tilde{\theta}_{2} \to \infty$;
    \item
    \label{illcase3} {the} horizontal line discontinuous in $\tilde{x}=0$ $\tilde{y} = q+ q^*\mathbbm{1}_{\tilde{x}>0}$, {when} $\tilde{\theta}_{0} = q$,
    $\tilde{\theta}_{1} = q^*\tilde{\theta}_{2}$, and
    $\tilde{\theta}_{2} \to 0$.    
\end{enumerate}
\end{proposition}
\begin{proof}[Proof of Proposition 1.]
    First note that the response curves are continuous functions of the parameters. This fact implies that the possible limiting curves that are not branches of hyperbola on $\mathcal{X}$ may only be found for the parameters that tend to the border of their domain. 
    Clearly, these curves will be monotone nondecreasing.
\begin{description} 
    \item[$\tilde{\theta}_{0}$:] for $\tilde{\theta}_{0}$ that diverges, no bounded curves are possible. 
    \item[$\tilde{\theta}_{1}$:] for $\tilde{\theta}_{1} \to\infty $, the class of admissible bounded increasing limiting curves arises when $\tilde{\theta}_{1} = O(\tilde{\theta}_{2})$ [limit class \ref{illcase1} and limit class \ref{illcase2} when $\tilde{\theta}_{1} = o(\tilde{\theta}_{2})$]. When $\tilde{\theta}_{1} \to 0$ and $\tilde{\theta}_{2}$ bounded away from $0$, the model is not identifiable as $\tilde{\theta}_{2}$ is meaningless (limit class \ref{illcase2}).
    \item[$\tilde{\theta}_{2}$:] for $\tilde{\theta}_{2} \to\infty$ and $\tilde{\theta}_{1}$ bounded from above, the model is not identifiable as $\tilde{\theta}_{1}$ is meaningless (limit class \ref{illcase2}). For $\tilde{\theta}_{2} \to 0$, the class of admissible bounded increasing limiting curves arises when $\tilde{\theta}_{1} = O(\tilde{\theta}_{2})$ [limit class \ref{illcase3} and limit class \ref{illcase2} when $\tilde{\theta}_{1} = o(\tilde{\theta}_{2})$]. \qedhere
\end{description}%
\end{proof}%

\begin{remark}
    Let us note that the three limit cases in Proposition~\ref{rem:cpt} correspond to common models in practice. In case 1, the response mean depends linearly  on the dose, in case 2 the response mean is independent on the dose and case 3 corresponds to the typical model for studying homeopathic therapies (where there is a dose-effect only after a null threshold $\tilde{x}=0$ or $x=a$).  
\end{remark}

\section{Maximum Likelihood Estimator}
 \label{sect:MLE}

In general, for a design \eqref{xi} with $M$ support points, the MLE can be found from the sufficient statistics 
$\bar{y}_{i} = (y_{i,1}+\cdots+y_{i,n_i})/n_i$ (the mean observed response value for each design point $x_i$, $i=1, \ldots M$):
\begin{equation}\label{eq:MLEtoBeMin}
 \hat{\btheta}_{ML} 
 =\arg\min_{\btheta} \sum_{i=1}^M n_i [\bar{y}_{i} - \eta(x_i,\btheta)]^2    
\end{equation}
\citep[see][]{Flou:May:Toma2021}. 
Figure~\ref{fig:ConcaveShape_points_curve} shows the MLE Emax response mean curve that fits $(x_i,\bar{y}_{i}), i= 1,2,3,$.

\begin{figure}[htbp]
    \centering
    \includegraphics[width = .75\linewidth]{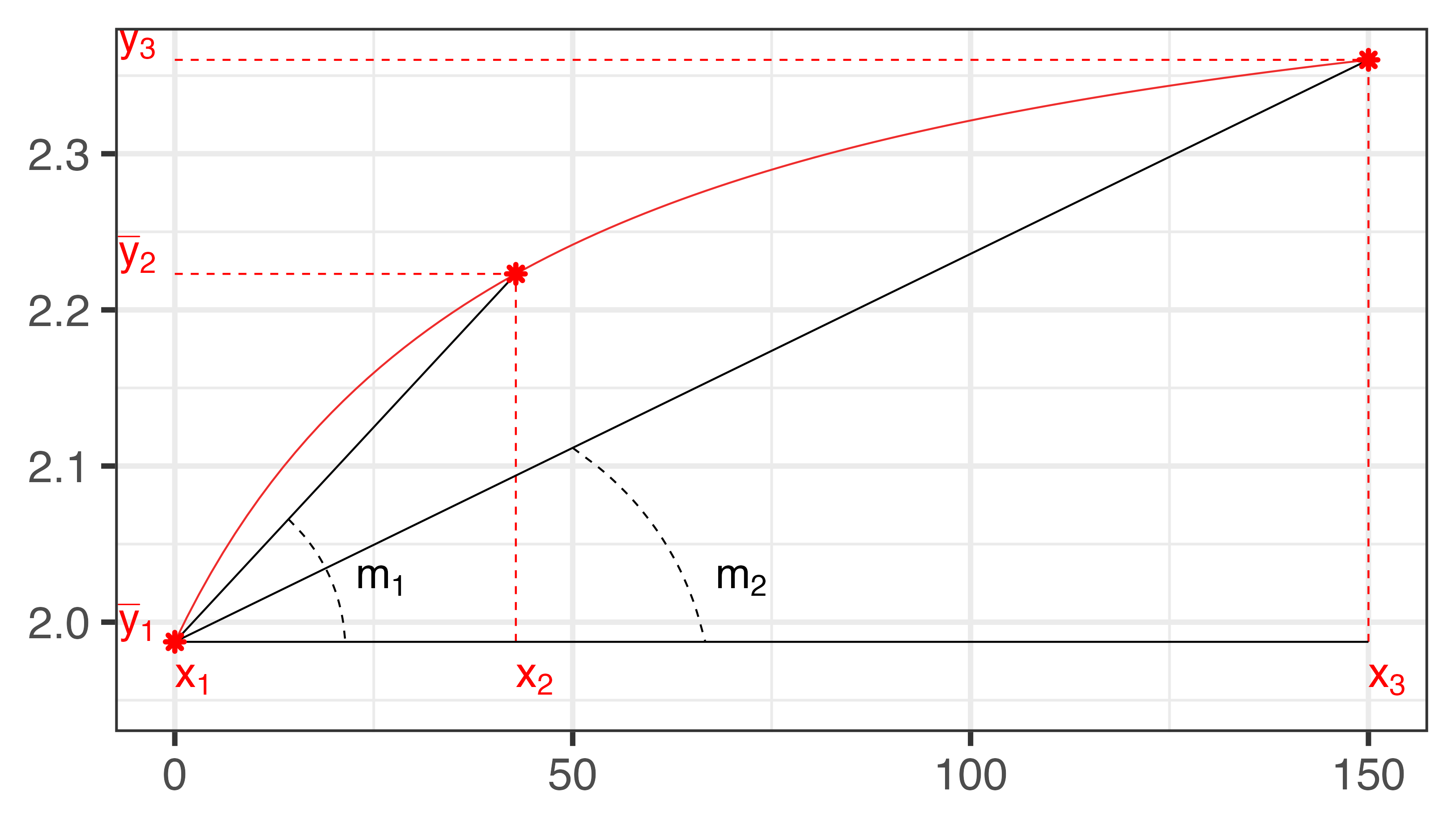}
    \caption{Example of points $(x_i,\bar{y}_{i}), i= 1,2,3,$ that have an increasing concave shape. The red curve is the Emax response mean that fits them.}
    \label{fig:ConcaveShape_points_curve}
\end{figure}

The next lemma states an important geometric result: an Emax model cannot be identified with a MLE when the best response fitting curve is given by a limit  in Proposition~\ref{rem:cpt}.
However, data sets like the one shown in Figure~\ref{fig:ConcaveShape_badCase} may occur.

\begin{figure}[htbp]
    \centering
  \includegraphics[width=.75\linewidth]{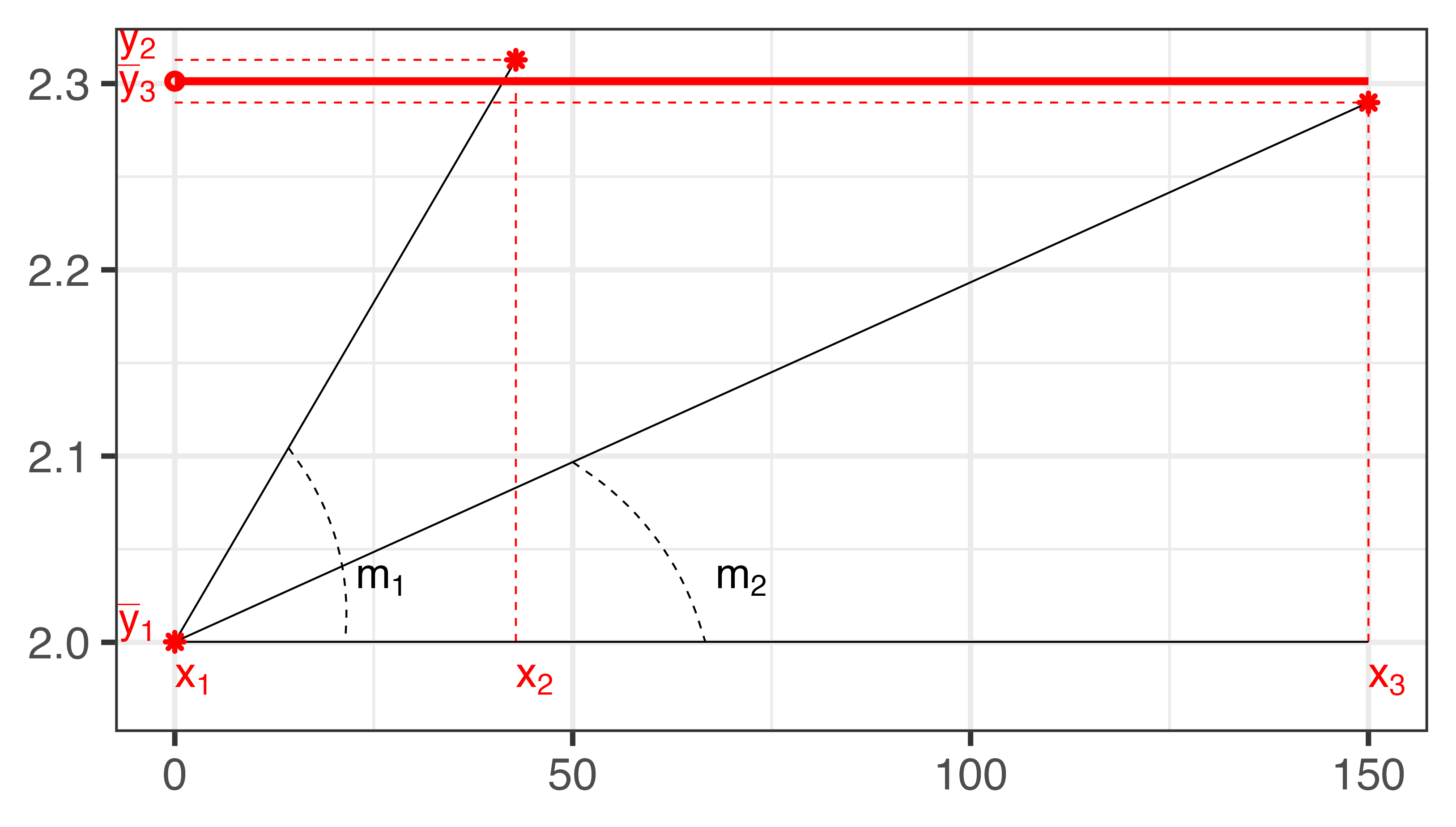}
    \caption{An example of three points $(x_i,\bar{y}_{i}), i= 1,2,3,$ that are concave, but non-increasing, are plotted as red asterisks. An example of the family of discontinuous curve comprising case \ref{illcase3} of Proposition \ref{rem:cpt}, which satisfies \eqref{eq:etaOutperfEmax}, is graphed in red.}
    \label{fig:ConcaveShape_badCase}
\end{figure}

\begin{lemma}\label{lem:limitMLE}
    If there exists a function $\eta$ in the limit class of Proposition~\ref{rem:cpt} that fits the data better than any Emax model, i.e.,  
\begin{equation}\label{eq:etaOutperfEmax}
\text{if for any }\btheta \qquad \sum_{i=1}^M n_i [\bar{y}_{i} - \eta(x_i)]^2 < \sum_{i=1}^M n_i [\bar{y}_{i} - \eta(x_i,\btheta)]^2,
\end{equation}
then the MLE does not exist. 
\end{lemma}

\begin{proof}[Proof of Lemma~\ref{lem:limitMLE}.]
For any $\epsilon > 0$, there exists $\btheta$ such that 
$$
0 <  \sum_{i=1}^M n_i [\bar{y}_{i} - \eta(x_i)]^2 - \sum_{i=1}^M n_i [\bar{y}_{i} - \eta(x_i,\btheta)]^2 \leq \epsilon;
$$
and hence the minimum of \eqref{eq:MLEtoBeMin} cannot be reached.
\end{proof}
Note that the probability of the non-existence of an MLE for an Emax model is always positive, regardless of the design, as the probability of \eqref{eq:etaOutperfEmax} is non-zero in any Emax model. Clearly, this probability approaches zero as the sample size increases or the variance of the Gaussian noise diminishes. Classifying the geometrical situations where \eqref{eq:etaOutperfEmax} holds is not a trivial task.
In this paper, it is solved for a 3-point design, which can serve as a benchmark and a guide for other designs. 
For designs with more than three points, the situation becomes more complex and is left for future research. The key issue is that, given a design with two consecutive internal points, 
$x_*$ and $x^*$ with $n_*$ and $n^*$ experimental units, it is possible to construct a new design by merging these two points into one with 
$n_*+n^*$ experimental units. Which design performs better in terms of the probability of \eqref{eq:etaOutperfEmax}? The answer is not straightforward due to the trade-off between exploring possible shapes by using more points and the uncertainty caused by reducing the number of experimental units per point. %

From this point onwards, we focus on a 3-point design
\begin{equation*}
     \xi
=\left\{
\begin{matrix}
x_1 & x_2 & x_3 \\
\omega_1 & \omega_2 & \omega_3
\end{matrix}
 \right\},
\end{equation*}
where  $x_1=a$ and $x_3=b$. We characterize the geometrical situations in which the MLE cannot be computed.   %, $\omega_i=n_i/n$. questo lo abbiamo gi‚àö‚Ä† detto nell'intro generale
Note that the D-optimal design \eqref{eq:D-opt}, in particular, is a equally supported 3-point design,  where 
 $x_2=x^*(\theta_2)$  given by Equation~\eqref{eq:D_opt_dose}.
The following remark highlights the special role of $x_2=x^*(\theta_2)$ for the model identification.   

\bigskip

\begin{remark}\label{rem:optChoiceXc}
It follows from Equation~\eqref{eq:D_opt_dose} that  $x^*(\theta_2) \in (a, (a+b)/2)$ since $\theta_2>-a$.
This helps the MLE to exist by taking Case 1 under control (see Figure~\ref{fig:ProbWP}).
Actually, the D-optimal design $ \xi^*$ increases the likelihood of  MLE existence.
%If we try to go forward in this path, 
The fact that the two extreme points of $\xi^*$ lie at the boundary of $\mathcal{X}$ is not surprising, but
it is interesting to note that $x^*(\theta_2)$ is the point where we observe the average value of the minimum and maximum mean responses, i.e. 
$\eta\left(x^*(\theta_2),\btheta\right)=[\eta(a,\btheta)+\eta(b,\btheta)]/2$.
This property maximizes the probability that $\bar{y}_1 < \bar{y}_2 < \bar{y}_3$, and this explains why the D-optimal design (by balancing the two cases given in Figure~\ref{fig:ProbWP})  reduces the probability of the MLE non-existence.
\end{remark}

\begin{figure}[htbp]
    \centering
    \includegraphics[width=0.8\linewidth]{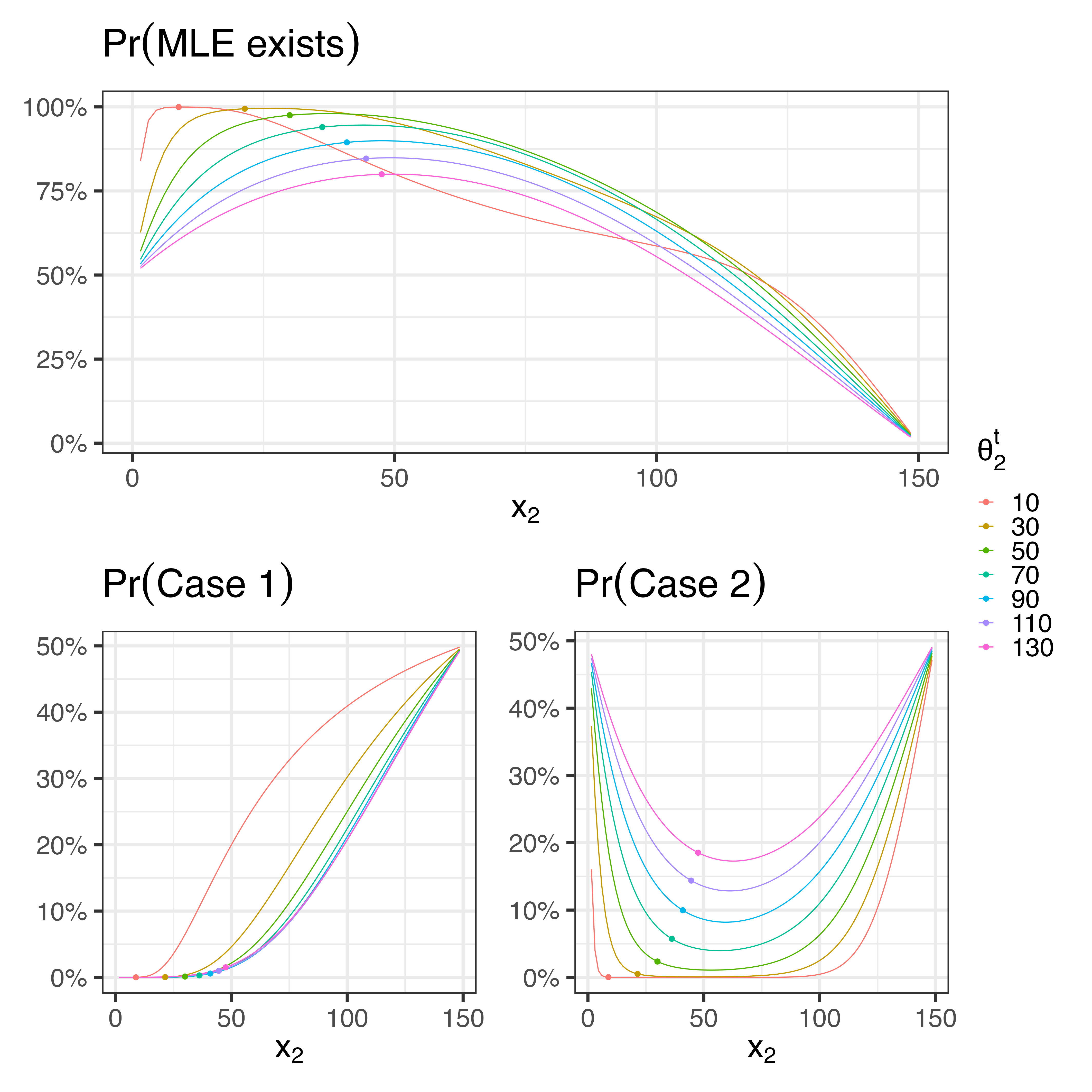}
        \caption{Probability that  the MLE exists [top figure, see Remark \ref{rem:2}] and that the MLE does not due to Case 1 and Case 2 [bottom left and right figures, respectively; see Remarks \ref{rem:3} and \ref{rem:4}], as a function of the central point $x_2$ for different values of $\theta_2^t$, where $\theta_0^t=2$, $\theta_1^t=0.467$, $a=0.001$, $b=150$, $\sigma=0.1$, $n_i=6$ (with $i=1,2,3$). The dot on the curves corresponds to  the D-optimal central point $x^*(\theta_2^t)$.}
        \label{fig:ProbWP}
\end{figure}

In what follows, we show that if the three points $(x_i,\bar{y}_{i}), i= 1,2,3,$ have an increasing concave shape (that is, the three points lay on an increasing concave curve as in Figure~\ref{fig:ConcaveShape_points_curve}), then there exists a unique analytic solution of the MLE. Otherwise, the existence %or the uniqueness 
of the MLE fails and we  provide the probability of observing data in the different adverse situations (see Figure~\ref{fig:ProbWP}).

\subsection{An Analytic Solution of the MLE for Data with Increasing Concave Shape}%\label{sub:analytic_sol}
The mathematical condition for the data having an increasing concave shape  is given by $\bar{y}_{1} < \bar{y}_{2} < \bar{y}_{3}$ and $m_1 > m_2$, where
\[
m_1 = \frac{\bar{y}_{2}- \bar{y}_{1} }{ x_{2} - x_{1}}, \qquad
m_2 = \frac{\bar{y}_{3}- \bar{y}_{1} }{ x_{3} - x_{1}} .
\]
Figure~\ref{fig:ConcaveShape_points_curve} displays a geometric {visualization} of this condition.

The following results provide the  analytic solution of the {unique MLE} for the reparametrization given in \eqref{eq:repEMAX} and for the model parameterization in \eqref{eq:E_max}. % given in \eqref{eq:norm_tilde} and for the parameter 

\smallskip

\begin{theorem}\label{theo:MLE}
Under the Emax model and a three points design, if the data have increasing concave shape then the MLE of $\tilde{\btheta}$  is given by
\begin{equation*}%\label{eqs:MLEtilde}
\begin{aligned}
\tilde{\theta}_{ML,0} 
& = \bar{y}_{1} 
\\
\tilde{\theta}_{ML,1} & = 
% m_2 (  \tilde{\theta}_{ML,2} + \tilde{x}_2   )
% \frac{ m_1 m_2 }{m_1-m_2} ({x}_3  -{x}_2  )
% =
 \frac{ m_1 m_2 }{m_1-m_2} (\tilde{x}_3  -\tilde{x}_2  )
\\
\tilde{\theta}_{ML,2} & = 
% \frac{m_2 \tilde{x}_{2} -m_1 \tilde{x}_{1} }{m_1-m_2} =
\frac{ \bar{y}_{3}  - \bar{y}_{2}  }{m_1-m_2} 
\end{aligned}
\end{equation*}
\end{theorem}
\begin{proof}[Proof of Theorem~\ref{theo:MLE}.]
Equation~\eqref{eq:MLEtoBeMin} can be rewritten in $\tilde{\btheta}$ and will be referenced in this parameterization during this proof.
From Equation~\eqref{eq:MLEtoBeMin}, with the new coordinates $\tilde{x} = x-a$, 
\begin{subequations}\label{eqr:All}
  \begin{align}
\label{eqr:0} %\tag{A0}
\bar{y}_{1} &=  %\eta(\tilde{x}_1,\tilde{\btheta}_{ML} )=
\tilde{\theta}_{ML,0} + \tilde{\theta}_{ML,1}\, \frac{\tilde{x}_{1}}{\tilde{x}_{1}+\tilde{\theta}_{ML,2}}
% = \tilde{\theta}_{ML,0} 
\\ \label{eqr:1} %\tag{A1}
\bar{y}_{2} &= %\eta(\tilde{x}_2,\tilde{\btheta}_{ML} )=
\tilde{\theta}_{ML,0} + \tilde{\theta}_{ML,1}\, \frac{\tilde{x}_{2}}{\tilde{x}_{2}+\tilde{\theta}_{ML,2}}
\\ \label{eqr:2} %\tag{A2}
\bar{y}_{3} &= %\eta(\tilde{x}_3,\tilde{\btheta}_{ML} )=
\tilde{\theta}_{ML,0} + \tilde{\theta}_{ML,1}\, \frac{\tilde{x}_{3}}{\tilde{x}_{3}+\tilde{\theta}_{ML,2}}.  
\end{align}
\end{subequations}
If the system of the three equations $\bar{y}_{i} = \eta(x_i,\tilde{\btheta} )$ $i=1,2,3$, is solved for a unique $\tilde{\btheta}$, then this solution must be a MLE. 
Recalling that $\tilde{x}_1 = 0$, $m_1= (\bar{y}_{2}-\bar{y}_{1})/\tilde{x}_2$ and $m_2= (\bar{y}_{3}-\bar{y}_{1})/\tilde{x}_3$, then the system \eqref{eqr:All} is equivalent to
\begin{subequations}
  \begin{align*}
\tag{\ref{eqr:0}}
\bar{y}_{1} &= 
\tilde{\theta}_{ML,0} 
\\ \tag{$\ref{eqr:2}-\ref{eqr:0}$}
\bar{y}_{3} - \bar{y}_{1} &= 
\tilde{\theta}_{ML,1}\, \frac{\tilde{x}_{3}}{\tilde{x}_{3}+\tilde{\theta}_{ML,2}}
\\ \tag{$\frac{\ref{eqr:2}-\ref{eqr:0}}{\ref{eqr:1}-\ref{eqr:0}}$}
\frac{m_2}{m_1}
 &= 
 \frac{\tilde{x}_{2}+\tilde{\theta}_{ML,2}}{\tilde{x}_{3}+\tilde{\theta}_{ML,2}}
= 1 - 
 \frac{\tilde{x}_{3}-\tilde{x}_{2}}{\tilde{x}_{3}+\tilde{\theta}_{ML,2}}
\end{align*}
\end{subequations}
and the thesis follows. % The following result is hence a simple consequence of \eqref{eq:norm_tilde}.   
\end{proof}

\begin{corollary}\label{cor:MLE}
Under the conditions of Theorem~\ref{theo:MLE}, the MLE of ${\btheta}$ is
\begin{align*}
\hat{\theta}_{ML,0} 
& = \bar{y}_{1} 
-a 
\frac{
m_1m_2(b-x_2)}{
(\bar{y}_{3}  - \bar{y}_{2}) - a (m_1-m_2)}
\\
\hat{\theta}_{ML,1} 
& = 
%(\bar{y}_{3}  - \bar{y}_{1} ) 
%\Big( 1 + \frac{ \bar{y}_{3}  - \bar{y}_{2}  }{m_1-m_2} \frac{1}{b-a} \Big)
\frac{ m_1 m_2 }{m_1-m_2} (b-x_2) 
+ a 
\frac{
m_1m_2(b-x_2)}{
(\bar{y}_{3}  - \bar{y}_{2}) - a (m_1-m_2)}
%\Big( 1 + \frac{a}{ \frac{ \bar{y}_{3}  - \bar{y}_{2}  }{m_1-m_2} - a } \Big)
\\
\hat{\theta}_{ML,2} 
& =
\frac{ \bar{y}_{3}  - \bar{y}_{2}  }{m_1-m_2} - a.
\end{align*}
\end{corollary}
\begin{proof}[Proof of Corollary~\ref{cor:MLE}.]
Express \eqref{eq:norm_tilde} in terms of $\btheta$ and apply Theorem~\ref{theo:MLE}.
\end{proof}

\begin{remark}
\label{rem:2}
The mathematical condition of increasing concave shape for the data can be rewritten as a linear inequality $A\bar{\mathbf{y}}  < \mathbf{0}$, with
\begin{equation*}
A = 
\begin{pmatrix}
\Big(\tfrac{1}{x_2-x_1} -\tfrac{1}{x_3-x_1}\Big) & -\tfrac{1}{x_2-x_1}  & \tfrac{1}{x_3-x_1} \\
    1 & -1\phantom{+} & \phantom{+}0\phantom{+} \\
    0 & \phantom{+}1\phantom{+} & -1\phantom{+}    
\end{pmatrix} 
,
\qquad
\bar{\mathbf{y}} =
\begin{pmatrix}
    \bar{y}_{1} \\ \bar{y}_{2} \\ \bar{y}_{3} 
\end{pmatrix}.    
\end{equation*}
Since $\bar{y}_{i} \sim N( \eta(x_i,\btheta), {\sigma^2}/{n_i})$ are independent, we are able to compute numerically the probability of $P(A\bar{\mathbf{y}}  < \mathbf{0})$. The top plot in Figure~\ref{fig:ProbWP} displays the probability that the MLE exists for various values of the central support point $x_2$.
\end{remark}
\bigskip

We have proved that, for data with increasing concave shape, the graph of $\eta(\cdot,\hat{\btheta}_{ML})$ contains the three points $(x_i,\bar{y}_{i})$, $i = 1,2,3$ because of system \eqref{eqr:All}. This will not be the case for other shapes; when \eqref{eqr:All} is not satisfied, we will always find a function $\eta$ in the limit class of Proposition~\ref{rem:cpt} such that \eqref{eq:etaOutperfEmax} holds for three point designs. Then the MLE will not exist by Lemma~\ref{lem:limitMLE}. 

If we include the limit case in the model  when there is no effect of the drug, that is, a constant response is obtained with $\theta_1=0$, then the model is not identifiable: equation \eqref{eq:MLEtoBeMin} may have more than one minimum. Another limit case, which is also not identifiable, is the linear increasing model (see \ref{illcase2} in Proposition~\ref{rem:cpt}), that can be included only by a reparametrization.

\subsection{Cases for which the MLE Does Not Exist}
\label{sec:no MLE}

Two data configurations in which the data do not exhibit an increasing concave shape lead to non-existent MLEs: 
\begin{itemize}
    \item[] {\bf Case 1: } the data exhibit a concave shape ($m_1>m_2$), but the three means $\bar{y}_{i}$, $i = 1,2,3$ are not increasing (see Figure~\ref{fig:firstTommasi});
    \item[] {\bf Case 2: } the three data points have convex shape ($m_1\leq m_2$), see Figure~\ref{fig:secondTommasi}.
\end{itemize}

\begin{figure}[htbp]
    \centering    \includegraphics[width=0.75\linewidth]{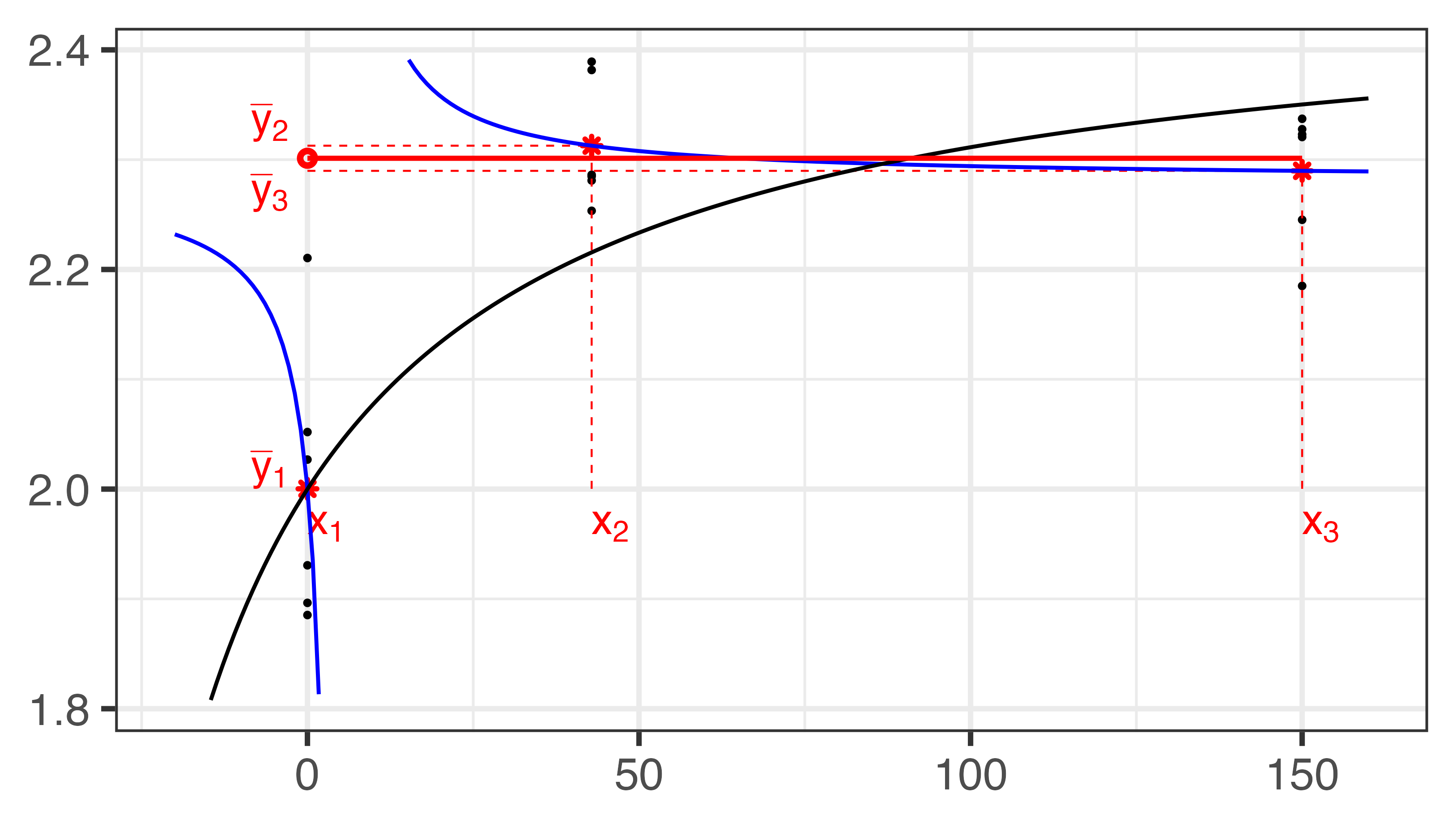}
    \caption{Case 1: data with non-increasing concave shape. The black line represents the true Emax model that generates the sample data (black dots). The red star points  are the sample means ($\bar y_i$) at the experimental points $x_i$, ($i=1,2,3$).  The blue curve displays the hyperbole that fits the three points $(x_i,\bar y_i)$, $i=1,2,3$ (i.e.\ model \eqref{eq:E_max} without any parametric constraint).  The discontinuous curve given in \eqref{eq:etaInSub1a}, which satisfies \eqref{eq:etaOutperfEmax}, is plotted in red.}
    \label{fig:firstTommasi}
%\end{figure}
%
%\begin{figure}[htbp]
    \centering
    \includegraphics[width=0.75\linewidth]{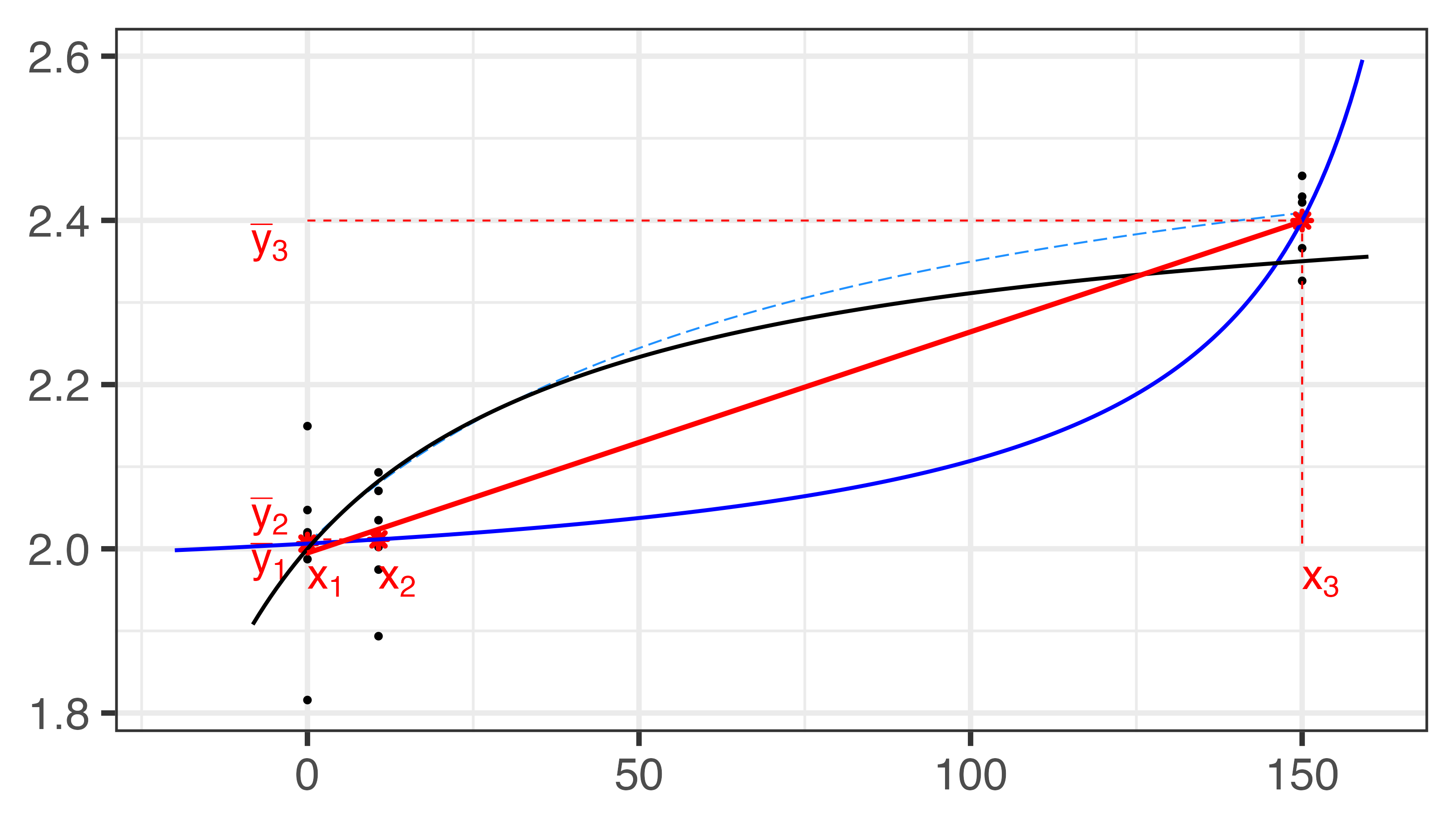}
    \caption{Case 2: data with convex shape. The black line represents the true Emax model that generates the sample data (black dots). The red star points  are the sample means ($\bar y_i$) at the experimental points $x_i$, ($i=1,2,3$).  The blue curve displays the hyperbole that fits the three points $(x_i,\bar y_i)$, $i=1,2,3$ (i.e.\ model \eqref{eq:E_max} without any parametric constraint).  The simple linear regression of the original data, which satisfies \eqref{eq:etaOutperfEmax}, is plotted in red. The light blue dashed-curve represents the Emax model that corresponds to the Firth-modified estimator.}
    \label{fig:secondTommasi}
\end{figure}

\noindent \textbf{Case 1: Data with non-increasing concave shape.}
In this case $m_1>m_2$, but the MLE existence requirement that $\bar{y}_{1} < \bar{y}_{2} < \bar{y}_{3}$ fails.

We start by showing that, in this case, $\bar{y}_{2} \geq \bar{y}_{3}$.
In fact, by contradiction, an assumption that $\bar{y}_{2} < \bar{y}_{3}$ implies $m_2>0$. Hence $m_1>0$ which implies that  $\bar{y}_{1} < \bar{y}_{2}$, which together with the assumption $\bar{y}_{2} < \bar{y}_{3}$ leads to the contradiction.

The fact is that $\bar{y}_{2} \geq \bar{y}_{3}$ is opposite to the expected response from the Emax model, where 
$\eta({x}_2,{\btheta}) < \eta({x}_3,{\btheta})$. 
Applying Lemma~\ref{lem:combin2} with $(x_*,y_*)=(x_2,\bar{y}_2)$, $(x^*,y^*)=(x_3,\bar{y}_3)$, $z_*=\eta({x}_2,{\btheta})$, $z^*=\eta({x}_3,{\btheta})$, $n_*=n_2$ and $n^*=n_3$ we obtain
\begin{multline}\label{eq:monotProperty}
n_2 \left[\bar{y}_{2} - \eta({x}_2,{\btheta})\right]^2 + n_3 \left[\bar{y}_{3} - \eta({x}_3,{\btheta})\right]^2 
> 
n_2 (\bar{y}_{2} - c)^2 + n_3 (\bar{y}_{3} - c)^2 
\\
\geq 
\min_d \sum_{i=2,3} n_i (\bar{y}_{i} - d )^2
%\\
%& 
= 
\sum_{i=2,3} n_i \big(\bar{y}_{i} - 
\bar{y}_{23}
\big)^2,
\end{multline}
where 
$$
\bar{y}_{23}=\frac{n_2\bar{y}_{2} + n_3\bar{y}_{3}}{n_2+n_3}.
$$ 
Equation \eqref{eq:monotProperty} states that any curve from the Emax model fits the two points $({x}_2,\bar{y}_{2})$ and $({x}_3,\bar{y}_{3})$ worse than the horizontal line $y = \bar{y}_{23}$. To complete \eqref{eq:MLEtoBeMin}, it remains to include in \eqref{eq:monotProperty} the term that relates to $(x_1,\bar{y}_{1})$. To do so, we consider two subcases that are distinguished by   the position of $\bar{y}_{1}$ with respect to $\bar{y}_{23}$. By Lemma~\ref{lem:limitMLE}, the MLE does not exist in either subcase. 
\begin{proposition}\label{prop:twosubcases_1}
Subcase $\bar{y}_{1} < \bar{y}_{23}$: when $\bar{y}_{1} < \bar{y}_{23}$, the function 
    \begin{equation}\label{eq:etaInSub1a}
\eta(x) =
\begin{cases}
    \bar{y}_{1} & \text{if $x = x_1=a$;}
    \\
    \bar{y}_{23} & \text{if $x \in (x_1$,b];}
\end{cases}
    \end{equation}
satisfies \eqref{eq:etaOutperfEmax} and $\eta$ belongs to the limit class \ref{illcase3} of Proposition~\ref{rem:cpt}.
\\
Subcase $\bar{y}_{1} \geq \bar{y}_{23}$: When $\bar{y}_{1} \geq \bar{y}_{23}$, the horizontal line
    \begin{equation}\label{eq:etaInSub1b}
 y = \bar y, \quad {\rm where}\; \bar y=\frac{n_1\bar{y}_{1} + n_2\bar{y}_{2} + n_3\bar{y}_{3}}{n_1+n_2+n_3},
    \end{equation}
satisfies \eqref{eq:etaOutperfEmax} and $\eta$ belongs to the limit class \ref{illcase2} of Proposition~\ref{rem:cpt}. 
\end{proposition}
\begin{proof}[Proof of Proposition~\ref{prop:twosubcases_1}]
Subcase $\bar{y}_{1} < \bar{y}_{23}$.
In this case, for any $\btheta$, by \eqref{eq:monotProperty} we obtain 
\begin{multline*}
\sum_{i=1}^3 n_i (\bar{y}_{i} - \eta(x_i,\btheta))^2 >
\sum_{i=2,3} n_i (\bar{y}_{i} - \eta(x_i,\btheta))^2
\\ 
> 
\sum_{i=2,3} n_i (\bar{y}_{i} - 
\bar{y}_{23}
)^2
=  
\sum_{i=1}^3 n_i (\bar{y}_{i} - \eta(x_i))^2 ,
\end{multline*}
where $\eta$ is defined in equation \eqref{eq:etaInSub1a}.

\medskip 

Subcase $\bar{y}_{1} \geq \bar{y}_{23}$.
%The first inequality of \eqref{eq:monotProperty} gives
Equation~\eqref{eq:nyz_*^*} with $y_*=\bar{y}_2$, $y^*=\bar{y}_3$, $z_*=\eta({x}_2,{\btheta})$, $z^*=\eta({x}_3,{\btheta})$, $n_*=n_2$ and $n^*=n_3$ gives
\begin{multline}\label{eq:3and2}
    \sum_{i=1,2,3} n_i (\bar{y}_{i} - \eta(x_i,\btheta))^2
     = 
    n_1 [\bar{y}_{1} - \eta(x_1,\btheta)]^2 +
    \sum_{i=2,3} n_i [\bar{y}_{i} - \eta(x_i,\btheta)]^2
\\
 > 
    n_1 [\bar{y}_{1} - \eta(x_1,\btheta)]^2 +
\sum_{i=2,3} n_i \Big(\bar{y}_{i} - 
\frac{n_2\eta(x_2,\btheta) + n_3\eta(x_3,\btheta)}{n_2+n_3}
\Big)^2.
\end{multline}
Let $\bar{\eta}_{23}(\btheta) = [n_2\eta(x_2,\btheta) + n_3\eta(x_3,\btheta)]/(n_2+n_3)$. The monotonicity of Emax model for $\theta_1>0$ implies that $\eta(x_1,\btheta) < \bar{\eta}_{23}(\btheta)$. We have to prove that, for any $\btheta$, there exists a constant $c = c(\theta)$ such that 
\begin{equation}\label{eq:SibPassagge}
\sum_{i=1,2,3} n_i (\bar{y}_{i} - \eta(x_i,\btheta))^2
 > 
\sum_{i=1,2,3} 
n_i (\bar{y}_{i} - c(\theta))^2.    
\end{equation}
We consider two subcases separately:
\begin{description}
    \item[Subcase $\bar{\eta}_{23}(\btheta) \leq \bar{y}_{1}$:]
    In this case $\eta(x_1,\btheta) < \bar{\eta}_{23}(\btheta) \leq \bar{y}_{1}$. Then $[\bar{y}_{1} - \eta(x_1,\btheta)]^2 > [\bar{y}_{1} - \bar{\eta}_{23}(\btheta)]^2$. By equation~\eqref{eq:3and2}, we obtain 
\begin{equation*}%\label{eq:monotProperty}
    \sum_{i=1,2,3} n_i [\bar{y}_{i} - \eta(x_i,\btheta)]^2
 > 
\sum_{i=1,2,3} n_i \Big(\bar{y}_{i} - \bar{\eta}_{23}(\btheta)
\Big)^2. 
\end{equation*} 
    \item[$Subcase\ \bar{\eta}_{23}(\btheta) > \bar{y}_{1}$:] The assumption of Subcase~1.b implies $ \bar{\eta}_{23}(\btheta) > \bar{y}_{1} \geq \bar{y}_{23}$.
    Since $f(c) = \sum_{i=2,3} n_i (\bar{y}_{i} - c )^2$ is a parabola with minimum at $\bar{y}_{23}$, then $f(\bar{\eta}_{23}(\btheta)) > f(\bar{y}_{1})$. Hence by \eqref{eq:3and2}
\begin{equation*}%\label{eq:monotProperty}
    \sum_{i=1,2,3} n_i [\bar{y}_{i} - \eta(x_i,\btheta)]^2
 > [\bar{y}_{1} - \eta(x_1,\btheta)]^2 +
f(\bar{y}_{1}) \geq \sum_{i=1,2,3} 
n_i (\bar{y}_{i} - \bar{y}_{1}
)^2. 
\end{equation*} 
\end{description}
By \eqref{eq:SibPassagge}
\begin{equation*}%\label{eq:monotProperty}
    \sum_{i=1,2,3} n_i [\bar{y}_{i} - \eta(x_i,\btheta)]^2
 > 
 \min_c
\sum_{i=1,2,3} 
n_i (\bar{y}_{i} - c)^2 
= 
\sum_{i=1,2,3} 
n_i (\bar{y}_{i} - \bar{y})^2 , %\qedhere
\end{equation*}
where $\bar{y}$ is the weighted mean of $\{\bar{y}_{i}$, $i=1,2,3\}$ and thus, Equation \eqref{eq:etaInSub1b} holds.
\end{proof}

\begin{remark} 
\label{rem:3}
The mathematical condition of non-increasing concave shape with $\bar{y}_{1} < \bar{y}_{23}$) (first subcase in Proposition~\ref{prop:twosubcases_1}) can be rewritten as a linear inequality $A\bar{\mathbf{y}}  < \mathbf{0}$, with
\[
A = 
\begin{pmatrix}
    \tfrac{1}{x_2-x_1} -\tfrac{1}{x_3-x_1} & -\tfrac{1}{x_2-x_1}  & \tfrac{1}{x_3-x_1} \\
    1 & -\frac{n_2}{n_2+n_3} & -\frac{n_3}{n_2+n_3} \\
    0 & -1 & 1    
\end{pmatrix} 
,
\qquad
\bar{\mathbf{y}} =
\begin{pmatrix}
    \bar{y}_{1} \\ \bar{y}_{2} \\ \bar{y}_{3} 
\end{pmatrix},
\]
while the mathematical condition of non-increasing concave shape with $\bar{y}_{1} \geq \bar{y}_{23}$ (second subcase in Proposition~\ref{prop:twosubcases_1}) can be rewritten as $A\bar{\mathbf{y}}  < \mathbf{0}$, with
\[
A = 
\begin{pmatrix}
    \tfrac{1}{x_2-x_1} -\tfrac{1}{x_3-x_1} & -\tfrac{1}{x_2-x_1}  & \tfrac{1}{x_3-x_1} \\
    -1 & +\frac{n_2}{n_2+n_3} & +\frac{n_3}{n_2+n_3} \\
    0 & -1 & 1    
\end{pmatrix} 
,
\qquad
\bar{\mathbf{y}} =
\begin{pmatrix}
    \bar{y}_{1} \\ \bar{y}_{2} \\ \bar{y}_{3} 
\end{pmatrix}.
\]
In both subcases, $\bar{y}_{i} \sim N( \eta(x_i,\btheta), {\sigma^2}/{n_i})$ are independent and the probabilities that observed data will belong to each subcase can be computed.
The bottom left plot in Figure~\ref{fig:ProbWP} displays the probability that the MLE fails (as a function of the central support point $x_2$) due to non-increasing concave data.
\end{remark}

\noindent \textbf{Case 2: Data with convex shape.}
Data with a convex shape ($m_1 \leq m_2$) is not  expected  from the Emax model whose  response curve is concave. This fact implies that any curve $\eta(x,\btheta)$ from the Emax model fits the three points $({x}_i,\bar{y}_{i})$ $i=1,2,3$ worse than a specific nondecreasing line, %
as is proved in Lemma~\ref{lem:cfrLines} in   Section~\ref{sec:aux_res} on auxillary results. We display the concept of Lemma~\ref{lem:cfrLines}  in Figure~\ref{fig:ConvexShape}. The  results for two subcases are described below. %
\begin{figure}[htb]
    \centering
    \includegraphics[width=.95\linewidth]{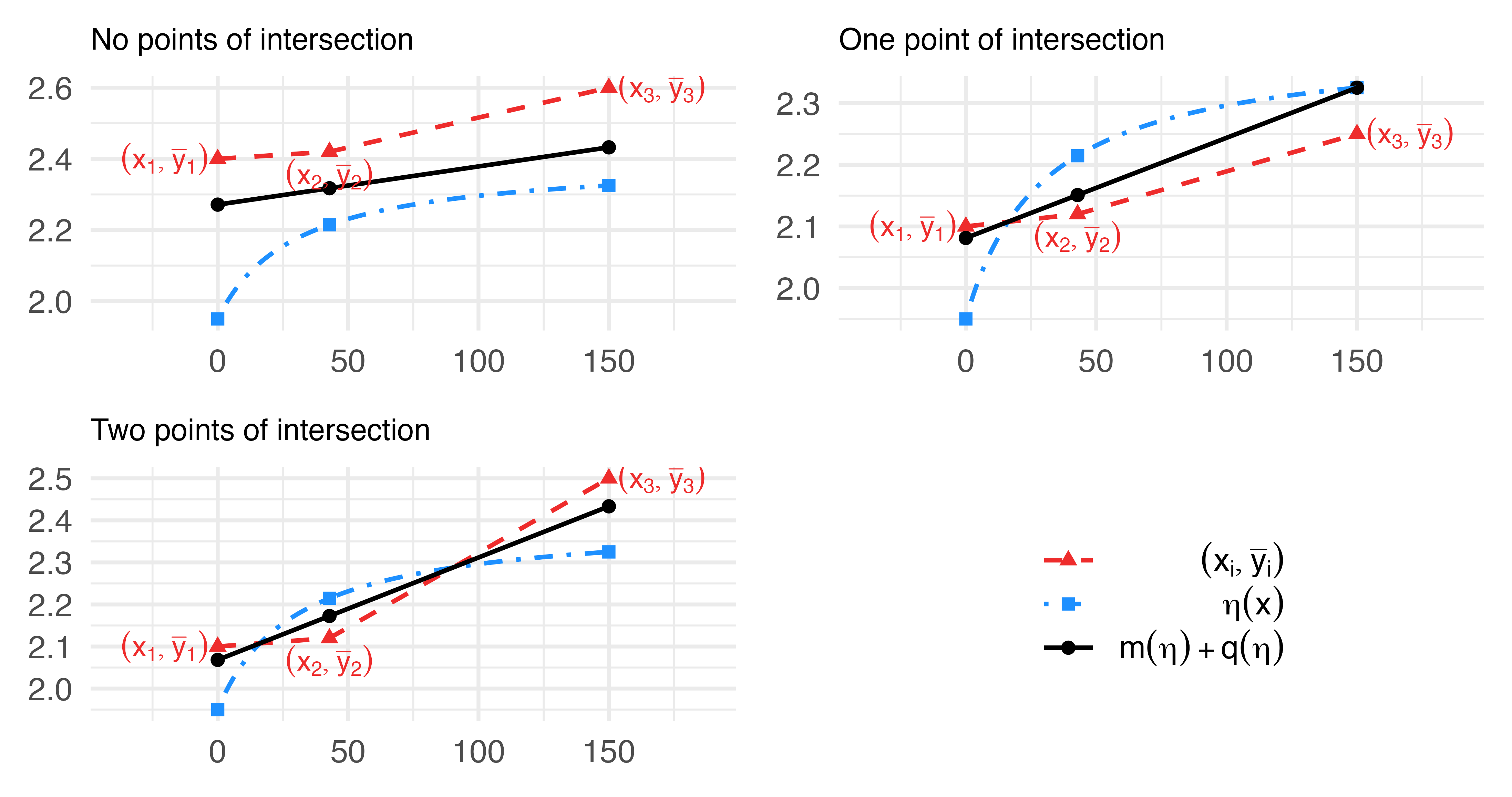}
    \caption{Examples of Case 2 non-existant MLEs providing the concept of the proof of  Lemma~\ref{lem:cfrLines}: the red triangles represent the data $\{(x_i,\bar{y}_i)$, $i=1,2,3\}$ that exhibit a convex shape. $\eta$ (blue dash-dotted line) is a concave increasing curve that might fit the data. The black straight line  fits the three red-triangles  better  than  $\eta$ (see Lemma~\ref{lem:cfrLines}).}
    \label{fig:ConvexShape}
\end{figure}

Let $\eta_0(x) = m_0 x+ q_0$ be the \emph{ordinary weighted simple linear regressor} of $(x_i,\bar{y}_i)$ $i=1,2,3$ -- that is the \emph{simple linear regressor} of the original data.  

\begin{description}
    \item[Subcase $m_0>0$:] Equation \eqref{eq:etaOutperfEmax} is satisfied with $\eta_0(x)$ since, by Lemma~\ref{lem:cfrLines},
    \begin{multline*}
       \sum_{i=1,2,3} n_i [\bar{y}_{i} - \eta(x_i,\btheta)]^2
 >
\sum_{i=1,2,3} n_i \big\{\bar{y}_{i} - [m(\eta)x_{i} + q(\eta)] \big\}^2 %\\
\geq
\sum_{i=1,2,3} n_i \big[\bar{y}_{i} - \eta_0(x_i)\big]^2 .
    \end{multline*}
    Moreover, $\eta_0$ belongs to  limit class \ref{illcase1} of Proposition~\ref{rem:cpt}.\\
    \item[Subcase $m_0\leq 0$:] Note that all the lines in \eqref{eq:ineq} have $m(\eta)\geq 0$ by Lemma~\ref{lem:cfrLines}, since $\eta(x,\btheta)$ is increasing, and hence the left derivative of $\eta(x,\btheta)$ in $x_3$ is positive. Then, by  Corollary~\ref{cor:combin}, for any $\btheta$ there exists a constant $c(\btheta)$ such that
\begin{equation*}%\label{eq:ineq}
    \sum_{i=1,2,3} n_i \big[ \bar{y}_{i} - (m(\eta)x_{i} + q(\eta)) \big ]^2
 \geq
\sum_{i=1,2,3} n_i \big[\bar{y}_{i} - c(\btheta) \big]^2 ,
\end{equation*}
so that, by Lemma~\ref{lem:cfrLines},
\begin{equation*}%\label{eq:monotProperty}
    \sum_{i=1,2,3} n_i [\bar{y}_{i} - \eta(x_i,\btheta)]^2
 > 
 \min_c
\sum_{i=1,2,3} 
n_i (\bar{y}_{i} - c)^2 
= 
\sum_{i=1,2,3} 
n_i (\bar{y}_{i} - \bar{y})^2 . %\qedhere
\end{equation*}
Equation \eqref{eq:etaOutperfEmax} holds with $y = \bar{y}$ which belongs to the limit class \ref{illcase2} of  Proposition~\ref{rem:cpt}.
\end{description}
For both subcases, by Lemma~\ref{lem:limitMLE}, the MLE does not exist.

\bigskip

\begin{remark}
\label{rem:4}
The mathematical condition of data with convex shape (Case 2) can be rewritten as a linear inequality $A\bar{\mathbf{y}}  < \mathbf{0}$, with

\[
P (m_1 \leq m_2) = P \Big(  
    \big(\tfrac{1}{x_2-x_1} -\tfrac{1}{x_3-x_1}\big)\bar{y}_{1} 
    -\tfrac{1}{x_2-x_1} \bar{y}_{2}  + \tfrac{1}{x_3-x_1} \bar{y}_{3} 
    \geq 0 \Big) 
\]
where $\bar{y}_{i} \sim N( \eta(x_i,\btheta), \sigma^2/n_i)$ are independent.
The bottom right plot in Figure~\ref{fig:ProbWP} displays the probability that the MLE fails (as a function of the central support point $x_2$) due to convex data.
\end{remark}

\section{Score Modification According to   Firth's Theory}
\label{sect:Firth}
To overcome the unlucky cases described in Section \ref{sec:no MLE},
 we provide  Firth's modification of the score function for the Emax model. It is known that the solution of  Firth's modified score sometimes provides a finite estimate when MLE fails 
\citep[see, for instance,][]{kosmidis2009biasb,kosmidis2009bias}.  
In developing explicit applications, both Firth and Kosmidis have focused on categorical response models  \citep[e.g.,][]{kosmidis2011multinomial,kosmidis2017multinomial,kosmidis2017multinomialb,kosmidis2021jeffreys,koll2021bias}. 
For exponential families in canonical parameterization,  \citet{firth1993bias} showed that his modified score equations are equivalent to maximizing the likelihood penalized with Jefferys invariant prior \citep{jefferys1946denomination}. 

\subsection{The Score Vector and Fisher Information}
In a regression model with normal errors  $\varepsilon_i=y_i-\eta(x_i,\btheta),i=1,\ldots,n$, the score vector $\mathbf{U}=\nabla \ln \mathcal{L}_n(\btheta; y,x)$ 
is
\begin{equation*}
\mathbf{U}=\dfrac{1}{\sigma^2}\sum_{i=1}^n [y_i-\eta(x_i,\btheta)] \: \nabla \eta(x_i,\btheta),  
\end{equation*}
 where  for the Emax model \eqref{eq:E_max} \citep[as shown by ][]{Dette:2010}
\begin{align*}
    &\nabla \eta(x_i,\btheta)=\left(1,\; \dfrac{x_i}{x_i+\theta_2},\; \dfrac{-\theta_1x_i}{(x_i+\theta_2)^2} \right)^T.
%& U^T=
 % \sum_{i=1}^n \dfrac{y_i-\eta(x_i,\theta)}{\sigma^2} \left( 1\quad  \dfrac{x_i}{x_i+\theta_2}\quad  \dfrac{-\theta_1 x_i}{(x_i+\theta_2)^2} \right).
\end{align*}
Now suppose the $t$-th component $U_t$ of the score vector $\mathbf{U}$ is adjusted to
\begin{align}\label{eq:U*}
   U_t^*=U_t+A_t, \quad t=1,\ldots,p, 
\end{align}
where the modification is
%$A_t=A_t^{(E)}$
%(see Firth). From matrix notation in Kosmidis we have that 
\begin{equation}\label{eq:At}
    A_t=\dfrac{1}{2}\trace\{I^{-1}(P_t+Q_t)\} \quad \textrm{with}
\end{equation}
\vskip.2cm
\begin{equation}\label{eq:PQ}
 P_t=E(\mathbf{U}\mathbf{U}^TU_t) \quad \mbox{and} \quad Q_t=E(-O\,U_t); 
\end{equation}
\vskip.3cm
\begin{lemma}\label{lemmaAt}
In a non-linear  model with normal errors, %$P_t=E(\mathbf{U}\mathbf{U}^TU_t)= 0$, so $A_t$ in \eqref{eq:At} reduces to
\begin{equation*}
    A_t=\dfrac{1}{2}\trace (I^{-1}Q_t).
\end{equation*}
\end{lemma}
\begin{proof}[Proof of Lemma~\ref{lemmaAt}.]
    In a non-linear  model with normal errors, $P_t=E(\mathbf{U}\mathbf{U}^TU_t)= 0$, so  the result follows directly from \eqref{eq:At}.
\end{proof}
Let $O$ and $I$ denote the observed  and expected (Fisher) information matrices on the $p$-dimensional vector $\theta$, respectively.
More specifically, the $ij$-th element of the observed matrix information $O$ is 
%\begin{equation*}
    $-{\partial^2}\ln \mathcal{L}_n(\btheta; y,x)/{\partial \theta_i \partial \theta_j}$, which  for the Emax model \eqref{eq:E_max}  is
 \begin{equation*}\label{eq:I}
    O=\dfrac{1}{\sigma^2}\sum _{i=1}^n \left(
\begin{array}{ccc}
 1 & { \frac{x_i}{{\theta_2}+x_i}} 
 & { \frac{-\text{$\theta_1$} x_i}{\left(\text{$\theta_2$}+x_i\right){}^2}} \\[8pt]
 { \frac{x_i}{\text{$\theta_2$}+x_i}} 
 & { \frac{x_i^2}{\left(\text{$\theta_2 $}+x_i\right){}^2}} 
 &
  \frac{x_i }{\left(\theta_2+x_i\right)^2}(y_i-\text{$\theta_0$}-2\text{$\theta_1$} \frac{x_i}{\theta_2+x_i})\\[8pt]
  \frac{- \theta_1 x_i}{\left(\theta_2+x_i\right)^2} \;\;
&   \frac{x_i }{\left(\theta_2+x_i\right)^2}(y_i-\text{$\theta_0$}-2 \text{$\theta_1$} \frac{x_i}{\theta_2+x_i})\;\;\;
&   \frac{3\theta_1^2 x_i^2}{\left(\theta_2+x_i\right)^4}\! -\! \frac{ 2\theta_1 \,x_i (y_i- \theta_0)}{(\theta_2+x_i)^3} \!
\end{array}
\right).
\end{equation*}
%{$O$ ‚àö¬Æ stata corretta all'elemento posto (2,3)!}
\vskip.5cm
\noindent The expected information matrix $I=E(O)$ is
\begin{equation*}
    I=\dfrac{1}{\sigma^2}
 \sum _{i=1}^n   \left(
\begin{array}{ccc}
 1 & { \dfrac{x_i}{\theta_2+x_i}} 
 & -{\text{$\theta_1$} \dfrac{x_i}{\left(\text{$\theta_2$}+x_i\right)^2}} \\[8pt]
 { \dfrac{x_i}{\theta_2+x_i}} 
 & { \dfrac{x_i^2}{\left(\theta_2 +x_i\right)^2}} 
 &  
  \dfrac{-\theta_1x_i^2}{\left(\theta_2+x_i\right)^3}\\[8pt]
  \dfrac{- \theta_1x_i}{\left(\theta_2+x_i\right)^2} 
&  \dfrac{-\theta_1 x_i^2}{\left(\theta_2+x_i\right)^3} 
&  \dfrac{\theta_1^2x_i^2}{\left(\theta_2+x_i\right)^4}
\end{array}
\right).
\end{equation*}

\subsection{The Score Modification $A_t$}

The following theorem applies  Firth's additive score modifications to the Emax model.
\begin{theorem} \label{th:At}
 For the Emax model, the additive score modifications $A_t$, $t=1,2,3$ are \begin{align*}
A_1 &=
\frac{1}{\theta_1 D}
( 
V_{1,1}\, M_{1,3}-{\rm Cov}_{12}\, M_{1,2}
)
\\
A_2 &= 
\frac{1}{\theta_1 D}
(V_{1,1}\,M_{2,4}-{\rm Cov}_{12}\,M_{2,3})
\\
A_3 &= 
-\frac{1}{D}
(V_{1,1}\,M_{2,5} -{\rm Cov}_{12}\,M_{2,4}).
\end{align*}
where, for $l_1 = 1,2$ and $l_2=1,\ldots,5$,\\
\begin{equation*}
M_{l_1,l_2} 
= 
{\rm E}_\xi \Big[
\frac{x_i^{l_1}}{(\theta_2+x_i)^{l_2}}\Big];
\qquad V_{l_1,l_2}
=
{\rm Var}_\xi \Big[
\frac{x_i^{l_1}}{(\theta_2+x_i)^{l_2}}\Big] =
M_{2l_1,2l_2}-M_{l_1,l_2}^2; \textrm{ and}
\end{equation*}

\begin{equation*}
{\rm Cov}_{12} 
= 
{\rm Cov}_\xi \Big[
\frac{x_i}{(\theta_2+x_i)},
\frac{x_i}{(\theta_2+x_i)^2}
\Big] = M_{2,3}-M_{1,1}\, M_{1,2};
\qquad 
D 
=
V_{1,1}\,V_{1,2}-{\rm Cov}_{12}^2.
\end{equation*}
 \end{theorem}
\begin{proof}[Proof of Theorem~\ref{th:At}.] 
From equation \eqref{eq:PQ}, after some computation, we  find that all items in the matrices $Q_t=\{Q_{t(i,j)}\}_{i,j=1,2,3}$, $t=1,2,3$, are null except those in positions $(2,3)$ and $(3,3)$.  Specifically,
\begin{equation}\label{eq:nullQ}
  Q_{t(i,j)}=0 \quad \mbox{for $(i,j)\neq (2,3)\, \mbox{or}\, (3,3)$ and $t=1,2,3$}
\end{equation}
while
\begin{align*}
Q_{1(2,3)} &= -\frac{1}{\sigma^2}\sum_{i=1}^n \frac{x_i}{(x_i+\theta_2)^2}
\!=\! -\frac{n}{\sigma^2}\, M_{1,2};
%= -\frac{n}{\sigma^2} {\rm E}_\xi\! \left[\! \frac{x_i}{(x_i+\theta_2)^2}\!\right], 
\\
Q_{1(3,3)}&= \frac{2\,\theta_1}{\sigma^2}
    \sum_{i=1}^n \!\frac{x_i}{(x_i+\theta_2)^3}
\!=\! \frac{2\,n\,\theta_1}{\sigma^2}\, M_{1,3};\\
Q_{2(2,3)}&=-\frac{1}{\sigma^2}\sum_{i=1}^n \dfrac{x_i^2}{(x_i+\theta_2)^3}
= -\frac{n}{\sigma^2} \, M_{2,3};  
\\
Q_{2(3,3)}&=\frac{2\,\theta_1}{\sigma^2} \sum_{i=1}^n \dfrac{x_i^2}{(x_i+\theta_2)^4}
=\frac{2\,n\,\theta_1}{\sigma^2} \, M_{2,4}; \\
Q_{3(2,3)}&=\dfrac{\theta_1}{\sigma^2}\sum_{i=1}^n \dfrac{x_i^2}{(x_i+\theta_2)^4} 
=\dfrac{n\,\theta_1}{\sigma^2}\, M_{2,4};
\\
Q_{3(3,3)}&= -\dfrac{2\,\theta_1^2}{\sigma^2}\sum_{i=1}^n \dfrac{x_i^2}{(x_i+\theta_2)^5}
=-\dfrac{2\,n\,\theta_1^2}{\sigma^2}  \, M_{2,5}.
\end{align*}
It follows from Lemma~\ref{lemmaAt}  for the Emax model,   that 
\begin{equation}
\label{eq:A}
    A_t=\dfrac{1}{2}\trace (I^{-1}Q_t)=I^{-1}_{(2,3)}\,Q_{t(2,3)}+\dfrac{1}{2}\,I^{-1}_{(3,3)}\,Q_{t(3,3)},
\end{equation}
where  $I_{(i,j)}^{-1}$ denotes the $(i,j)$-th element of the expected information matrix $I^{-1}$ and the second equality follows from \eqref{eq:nullQ}.
The computation of $A_t$, $t=1,2,3$, requires only the elements in positions $(2,3)$ and $(3,3)$ of the inverse of the Fisher information matrix. Therefore, we have partitioned $I$ as follows
\begin{equation}
I=\begin{pmatrix}
\mathcal{I}_{11} & \mathcal{I}_{12}\\ \mathcal{I}_{21} & \mathcal{I}_{22}
\end{pmatrix},
\label{eq:partitioned_matrix}
\end{equation}
where 
$$
\mathcal{I}_{11}=\dfrac{n}{\sigma^2},
\quad 
\mathcal{I}_{21}=\dfrac{1}{\sigma^2}
\begin{pmatrix}
\displaystyle \sum _{i=1}^n \dfrac{x_i}{\theta_2+x_i} \\[12pt] 
-\theta_1 \displaystyle\sum _{i=1}^n \dfrac{x_i}{\left(\theta_2+x_i\right)^2}
\end{pmatrix}
=
\dfrac{n}{\sigma^2}
\begin{pmatrix}
M_{1,1} \\
-\theta_1\, M_{1,2}
\end{pmatrix},
\quad  
\mathcal{I}_{12}=\mathcal{I}_{21}^T 
\; \; {\rm and}  
$$
$$
\mathcal{I}_{22}=\dfrac{1}{\sigma^2} 
\begin{pmatrix}
\displaystyle\sum _{i=1}^n \dfrac{x_i^2}{\left(\theta_2 +x_i\right)^2} & -\theta_1 
\displaystyle \sum _{i=1}^n \dfrac{x_i^2}{\left(\theta_2+x_i\right)^3}\\[12pt]
 -\theta_1 \displaystyle\sum _{i=1}^n \dfrac{x_i^2}{\left(\theta_2+x_i\right)^3} & \theta_1^2 \displaystyle\sum _{i=1}^n \dfrac{x_i^2}{\left(\theta_2+x_i\right)^4}
\end{pmatrix}
=
\dfrac{n}{\sigma^2} 
\begin{pmatrix}
M_{2,2} & -\theta_1 
M_{2,3}\\
 -\theta_1 M_{2,3} & \theta_1^2 M_{2,4}
\end{pmatrix}.
$$
From \eqref{eq:partitioned_matrix}, the following formula for $I^{-1}$ applies: 
\begin{equation*}
  I^{-1} =
  \begin{pmatrix}
  \mathcal{I}^{11}
  & \mathcal{I}^{12}
  \\[.1cm]
  \mathcal{I}^{21}
 & \mathcal{I}^{22}
  \end{pmatrix},
\end{equation*}
where $\mathcal{I}^{12}={\mathcal{I}^{21}}^T$,
\begin{align}
    \mathcal{I}^{11}&=
  \mathcal{I}_{11}^{-1}+\mathcal{I}_{11}^{-1} \mathcal{I}_{12}\, \mathcal{I}^{22} \,  \mathcal{I}_{21} \mathcal{I}_{11}^{-1},
    \notag %\label{eq:I_11}
    \\
    \mathcal{I}^{21}&= 
    - \mathcal{I}^{22}\, 
    \mathcal{I}_{21} \mathcal{I}_{11}^{-1},
    \notag %\label{eq:I_21}
    \\
    \mathcal{I}^{22}&=(\mathcal{I}_{22}-\mathcal{I}_{21} \mathcal{I}_{11}^{-1} \mathcal{I}_{12})^{-1}.
    \label{eq:I_22}
\end{align}
From \eqref{eq:I_22}, after some algebra, one obtains
\begin{equation*}
%\label{eq:G22}
\mathcal{I}^{22}=
\begin{pmatrix}
 I^{-1}_{(2,2)} & I^{-1}_{(2,3)}\\
 I^{-1}_{(3,2)} & I^{-1}_{(3,3)}
\end{pmatrix}
= 
\frac{\sigma^2}{n \, \theta_1^2\, D}
\begin{pmatrix}
\theta_1^2\, V_{1,2}
&
\theta_1\, {\rm Cov}_{12}
\\
\theta_1\, {\rm Cov}_{12}
&
V_{1,1}
\end{pmatrix}. 
\end{equation*}
Substituting the expressions 
\begin{equation*}
I^{-1}_{(2,3)} = 
\frac{\sigma^2 }
{n\,\theta_1  D}
\, 
{\rm Cov}_{12}
\quad {\rm and} \quad
I^{-1}_{(3,3)} = 
\frac{\sigma^2 }
{n\,\theta_1^2 D}
\,
V_{1,1}
\end{equation*}
  into $Q_{t(2,3)}$ and $Q_{t(3,3)}$ in \eqref{eq:A}, we obtain:
\begin{align*}
A_1 &=
\frac{1}{\theta_1 D}
(V_{1,1}\, M_{1,3}-\text{Cov}_{12}\, M_{1,2})
\\
A_2 &= 
\frac{1}{\theta_1 D}
(V_{1,1}\,M_{2,4}-\text{Cov}_{12}\,M_{2,3})
\\
A_3 &= 
-\frac{1}{D}
(V_{1,1}\,M_{2,5} -\text{Cov}_{12}\,M_{2,4}). %\qedhere
\end{align*}
\end{proof}
%where $D$ depends on $\theta_2$ and $\xi$,  as shown in \eqref{eq:D}.
Note that the modification $\mathbf{A}=(A_1,A_2,A_3)^T$ of the score vector depends on $\theta_1$ and $\theta_2$. Furthermore,  $\mathbf{A}$ is  a function of the  design $\xi$ which enables us to explore the impact of the design choice on the modified score $\mathbf{U}^*=(U_1^*,U_2^*,U_3^*)^T$. 

\begin{remark}
In the Emax model, the derivative matrix of $\mathbf{U}^*$ expressed in \eqref{eq:U*} is not symmetric, and hence there is no  penalized likelihood  corresponding to the modified score $U^*_t$.
\end{remark}

%%%%%%%%%%%%%%
\section{Firth's Modified Scores and a Design-Based Solution for Non-Existent MLEs}
\label{sec:simulation}
We begin by  exploring, via simulations, whether the score modification proposed by Firth overcome the non-existence of the MLE (in Cases 1 and 2) by providing admissible solutions of $\mathbf{U}^*=\mathbf{0}$.

To simulate the data we fix $a=0.001$, $b=150$, $\theta_0^t=2$, $\theta_1^t=0.467$, $\theta_2^t=50$, and $\sigma=0.1$ as in  \citet{Dette:2012}.  
We generate $6$ responses at the experimental conditions $x_1=a$, $x_2=x^*(\theta_2)$ and $x_3=b$,  and  compute the sample  means $\bar y_1$, $\bar y_2$ and $\bar y_3$. When these sample data do not have an increasing concave shape, the MLE cannot be computed and we try to compute the Firth's modified estimate.
We replicate the above computations $N=10000$ times with the goal of computing the proportion of times that:
\begin{itemize}
    \item the MLE exists;
    \item  the data belong to Case 1 and  Firth's modification succeeds in finding an admissible estimate;
    \item the data  belong to  Case 2 and  Firth's  modification succeeds in finding an admissible estimate.
\end{itemize}   
To explore the dependence of these proportions on the choice of guessed values of $\theta_2$, say $\theta_2^g$, we replicate the above simulation study for $\theta_2^g\in \{12.5, 25, 50, 75, 100 \}$.

Table \ref{tab:sim} displays the simulation results. 
From the second column we can immediately observe that the proportion of times that the MLE exists is consistent with the theoretical probability (described in Remark \ref{rem:2}) which is reported in parenthesis. The same consistency is displayed in columns 3 and 5, for Case 1 and Case 2, respectively.

\begin{table}[h]
% \tbl{Summary of simulation results}
\centering
{\begin{tabular}[t]{cllclc}
\hline
Nominal &\%   &\% &\% Firth's success  & \% & \& Firth's success \\
$\theta_2$ & MLE exists &  Case 1 & with Case 1 &  Case 2 &  with Case 2\\
\hline
\cellcolor{gray!6}{12.5} & \cellcolor{gray!6}{84.48 (84.82)} & \cellcolor{gray!6}{0.00 (0.00)} & \cellcolor{gray!6}{NA} & \cellcolor{gray!6}{15.52 (15.18)} & \cellcolor{gray!6}{100.00}\\
25 & 93.69 (93.74) & 0.00 (0.01) & NA & 6.31 (6.25) & 99.84\\
\cellcolor{gray!6}{50} & \cellcolor{gray!6}{97.50 (97.53)} & \cellcolor{gray!6}{0.10 (0.12)} & \cellcolor{gray!6}{0.00} & \cellcolor{gray!6}{2.40 (2.35)} & \cellcolor{gray!6}{99.58}\\
75 & 98.12 (98.01) & 0.44 (0.47) & 0.00 & 1.44 (1.53) & 100.00\\
\cellcolor{gray!6}{100} & \cellcolor{gray!6}{97.91 (97.77)} & \cellcolor{gray!6}{0.85 (0.98)} & \cellcolor{gray!6}{1.18} & \cellcolor{gray!6}{1.24 (1.25)} & \cellcolor{gray!6}{100.00}\\
\hline
\end{tabular}}
\caption{Percent times that the MLE exists (column 2) or does not (columns 3 and 5), with the theoretical probability  reported in parenthesis. Columns 4 and 6 display the proportion of times that Firth's correction succeeds in finding an admissible estimate in Cases 1 and 2, respectively. NA = Not Applicable.}
\label{tab:sim}
\end{table}
Case 1 instead remains unsolved, as shown in column 4:
there are few Case 1 problems, but almost all of them cannot be  solved by Firth's correction.  
In contrast, Firth's modified  score function virtually always provides  admissible estimates for data in Case 2 (see column 6).

Since Case 1 is not solved by Firth's correction, we propose an alternative that involves choosing the experimental point $x_2$ accordingly to an hypothesis test on $\theta_2$ in
 the next section.

%%%%%%%%%%%%%%%%%%%%
\subsection{Design Choice Based on a Hypothesis Test Provides Modified MLE solutions for Case 1 Data}
We construct a hypothesis test in which data  with sample response means that display a  non-increasing concave shape fall into the rejection region.

Consider the  hypothesis test:
\begin{align}
%\begin{cases}
 H_0&:\ \theta_2 \geq \theta_2^{g}\\
 H_1&:\ \theta_2 < \theta_2^{g}
%\end{cases}
\end{align}
where $\theta_2^{g}$ is a guessed value such that $\theta_2^{g}>-a$ and reject the null hypothesis whenever the data belongs to Case 1. 
The power function of this test, $\beta(\theta_2;x_2)$, is shown in Figure~\ref{fig:ProbWP}. It is a  function of both $\theta_2$ and the experimental condition $x_2$. Therefore, given a design point $x_2$, the significance level is $\alpha=\beta(\theta_2^g;x_2)$. See Figure \ref{fig:x2_choice} for a graphical representation of %$x_c$
$x_2$
as a function of $\alpha$ and $\theta_2^g$.

\begin{figure}[h]
    \centering
\includegraphics[width=0.75\linewidth]{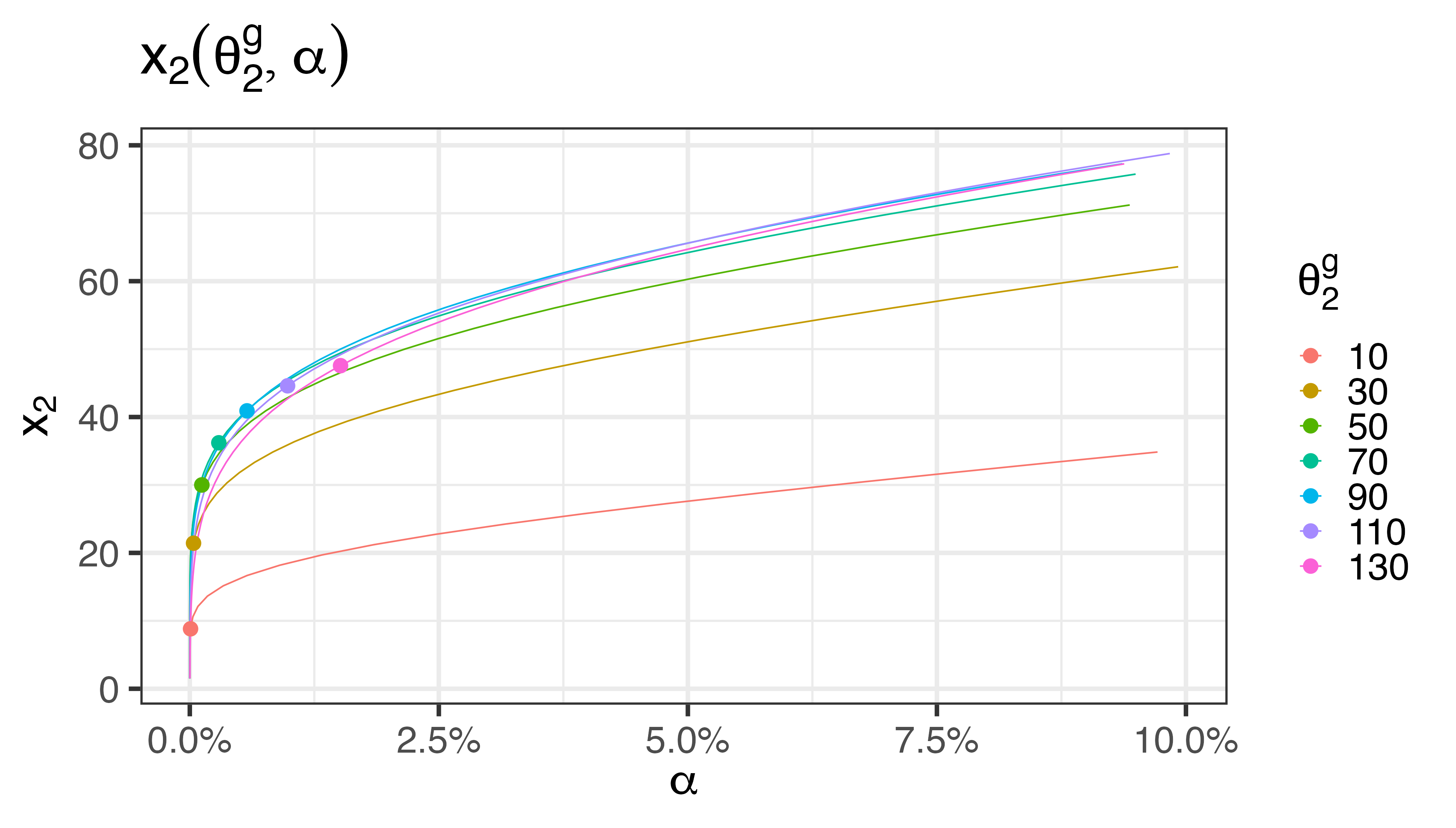}
    \caption{Central experimental point $x_2$ as a function of $\alpha$, for different $\theta_2^g$ (the graph has been obtained by changing $x$ and $y$ in Figure~\ref{fig:ProbWP}). The dots on the curves represent the central points of the D-optimal designs $x_2^*(\theta_2^g)$ given by Equation \eqref{eq:D_opt_dose}.}
    \label{fig:x2_choice}
\end{figure}

Reversing the role between $x_2$ and $\alpha$, this last statement can be used by an experimenter to properly choose the value of $x_2$ that guarantees a specific significance level $\alpha$. 
In other words, by solving for $x_2$ the equation $\alpha=\beta(\theta_2^g;x_2)$, one obtains the central design point 
%$x_c=
$x_2(\theta_2^g,\alpha)$ and by solving this equation with a small $\alpha$ guarantees a small  probability of observing a bad sample belonging to  Case 1. 

Let us observe that the D-optimal design values for  $x_2$ (the dots on the curves of Figure~\ref{fig:x2_choice}) correspond to very low $\alpha$-values.  Thus, if $H_0$ is true,  fixing a low value for $\alpha$ and taking the same proportion of observations at $a$, 
%$x_c$ 
$x_2$
and $b$,  should  protect from sample  responses that do not produce a MLE. 

If, despite choosing  a small $\alpha$, we observe a bad sample belonging to  Case 1,  $H_1$ is more likely than $H_0$. In this last case, we suggest  collecting additional  data at another experimental point $x_2(\theta_2^{(1)},\alpha)\in
(a,\ x_2(\theta_2^g,\alpha)), \textrm{ for some } \theta_2^{(1)},\ \theta_2^g$ informed by Figure~\ref{fig:x2_choice}.  
%%%%%%%%%%%%%%%%%%%%
\section{Practical Guidelines and Conclusions}\label{sect:concl}
Taking into account all the results presented in the previous sections,  we can provide some simple guidelines for gathering data that  produce (or are likely to produce) a finite estimate of the parameters of the Emax model. 

It is well known that the D-optimal design minimizes the generalized variance of the MLE but  other choices are also possible, for instance the A-optimality criterion minimizes the total variation of the MLE.
Herein, we suggest using  a locally D-optimal design, that it  is equally supported at three points, $x_1=a$,  $x_2=x^*(\theta_2^{g})\in \big(a,(a+b)/2\big)$  (see Remark~\ref{rem:optChoiceXc}) and $x_3=b$, where $\theta_2^{g}$ is a guessed value for $\theta_2$, because it guarantees (with a large probability) the existence of the MLE, as shown in Figure \ref{fig:ProbWP}.

Once the responses $y_{i1},\ldots,y_{in/3}$  (for $i=1,2,3$) have been observed at $x_1$, $x_2$ and $x_3$, respectively, we suggest to proceed as follows:
\begin{itemize}
    \item if the sample means $\bar y_i$ (for $i=1,2,3$) display an increasing concave shape, then the MLE can be computed (see Theorem~\ref{theo:MLE});
    \item if the sample means are in convex shape, one should  compute the Firth score modification that most likely leads to a finite parameter estimation (see Table~\ref{tab:sim});
    \item if the sample means are in non-increasing concave shape, then it is likely that $\theta_2^g$  has been chosen too high, therefore, no estimation is expected from Firth correction (see Table~\ref{tab:sim}). In this case, additional observations should be recorded at a new experimental point $x^*\big(\theta_2^{(1)}\big) \in \big(a,x^*(\theta_2^{g})\big)$, for some value $\theta_2^{(1)}<\theta_2^{g}$, as we have the statistical evidence that $\theta_2^t$ is significantly smaller than $\theta_2^{g}$.
\end{itemize}

{We believe that the theoretical contributions presented in this paper not only offer deep insights into the properties of the Emax model, but can also enhance significantly its estimation process in practical applications.}

%\section{Conclusions}\label{sec:conclude}
%%%%%%%%%%%%%%%%%%%%%%%%%%%%%%%%%%%%%%%%%%%%%%%%%%%%%%%%%%%%%%%%%%%%%%%%%%%%%%%%%%%%%%%%

\appendix

\section{Auxiliary results}\label{sec:aux_res}
In this section, to make the paper self contained, we prove some additional technical results that are useful for completeness.
\begin{lemma}\label{lem:cfrLines}
Let $(x_i,\bar{y}_{i}), i=1,2,3$ such that $x_1<x_2<x_3$ and assume
\[
m_1 = \frac{\bar{y}_{2}-\bar{y}_{1}}{x_2-x_1}
\leq 
\frac{\bar{y}_{3}-\bar{y}_{1}}{x_3-x_1}
= m_2.
\]
For any concave function $\eta:\mathcal{X}\to \mathbb{R}$, there exist a line $y = m(\eta) x + q(\eta)$ such that
\begin{equation}\label{eq:ineq}
    \sum_{i=1,2,3} n_i [\bar{y}_{i} - \eta(x_i)]^2
 \geq
\sum_{i=1,2,3} n_i \big\{\bar{y}_{i} - [m(\eta)x_{i} + q(\eta)] \big\}^2 ,
\end{equation}
with the equality being possible only when $\eta$ is a segment such that $\eta(x)\equiv m(\eta)x_{i} + q(\eta)$ on $\mathcal{X}$.\\
Moreover, $m(\eta)$ may be chosen to be not less than the left derivative of $\eta$ in $x_3$.
\end{lemma}
\begin{proof} [Proof of Lemma~\ref{lem:cfrLines}.]
Let $\zeta:\mathcal{X}\to\mathbb{R}$ be the convex function whose graph is obtained with the segments that link $(x_1,\bar{y}_1)$ with $(x_2,\bar{y}_2)$ and $(x_2,\bar{y}_2)$ with $(x_3,\bar{y}_3)$, i.e.,
\[
\zeta(x) = \sup\{ f:{\mathcal{X}}\to\mathbb{R} \text{ such that $f$ is convex and $f(x_i) \leq \bar{y}_i$, $i=1,2,3$}\}.
\]
Denote with $\gr(\eta)$ and $\gr(\zeta)$ the graphs of $\eta$ and $\zeta$, respectively.
The main idea of this proof is to find a line $y = mx+q$ that ``separates'' $\gr(\eta)$ from $\gr(\zeta)$, which  means that
\begin{equation}\label{eq:etaLineZeta}
\forall x \in {\mathcal{X}}, \qquad \eta(x) \leq mx+q \leq \zeta(x) \quad \mathrm{ or } \quad \zeta(x) \leq mx+q \leq \eta(x) .
\end{equation}
Condition \eqref{eq:etaLineZeta} is sufficient to state \eqref{eq:ineq}. Note that, if $\gr(\eta)$ is not a segment, then at least one of the two points $(x_1,\eta(x_1))$ or $(x_3,\eta(x_3))$ does not belong to the line $y=mx+q$, and hence \eqref{eq:ineq} is strict. We  prove \eqref{eq:etaLineZeta} for different cases.

\bigskip

\noindent\textbf{Case when $\eta$ and $\zeta$ are separated} (see Figure~\ref{fig:ConvexShape} top-left plot)\\
Assume that $\eta\leq\zeta$ on $\mathcal{X}$ or, vice versa, $\zeta\leq\eta$ on $\mathcal{X}$. Then the convex hulls of their graphs are also separated. The hyperplane separation theorem ensures the existence of a line $y(x) = mx+q$ that weakly separates the two closed hulls, so that, in this case a) $\forall x \in {\mathcal{X}},\eta(x) \leq mx+q \leq \zeta(x)$ or b) $\forall x \in {\mathcal{X}},\zeta(x) \leq mx+q \leq \eta(x)$. Denote by $L_{x_3}$ the left derivative of $\eta$ in $x_3$ and observe that it bounds from below any left and right derivatives of $\eta$, since $\eta$ is concave. Assume $m<L_{x_3}$. If a), then the line $y-(mx_3+q)=L_{x_3}(x-x_3)$ also separates $\eta$ and $\zeta$ as well as, in case b), the line $y-(mx_1+q)=L_{x_3}(x-x_1)$ does.

\bigskip

\noindent\textbf{Case when $\boldsymbol{\eta}$ and $\boldsymbol{\zeta}$ are not separated}
When $\eta$ and $\zeta$ are not separated, there exist two values $x_*,x^*\in{\mathcal{X}}$ such that $\eta(x_*)<\zeta(x_*)$ and $\eta(x^*)>\zeta(x^*)$. Bolzano's Theorem ensures the existence of an interior point $x_o \in (x_1,x_3)$ such that $\eta(x_o)=\zeta(x_o)$. Note that it is not possible for there to be more than two intersecting points between $\eta$ and $\zeta$ that are not separated and are concave and convex, respectively. Finally, we will find the separating line $y=mx+q$ by connecting two points of $\eta$, and then its slope cannot be less than the left derivative of $\eta$ in $x_3$.

\smallskip

\noindent\textbf{Subcase: $x_o\in(x_1,x_3)$ is the only point in common between the not separated $\boldsymbol{\eta}$ and $\boldsymbol{\zeta}$} (see Figure~\ref{fig:ConvexShape} top-right plot):\\
Depending on the reciprocal position of  $\eta$ and $\zeta$ we can have a) $\bar{y}_3<\eta(x_3)$ and $\bar{y}_1>\eta(x_1)$ or b) $\bar{y}_1<\eta(x_1)$ and $\bar{y}_3>\eta(x_3)$. If a) the line $y=mx+q$ connecting $(x_o,\eta(x_*o)$ and $(x_3,\eta(x_3))$ separates $\eta$ and $\zeta$. If b) the line $y=mx+q$ connecting $(x_o,\eta(x_o))$ and $(x_1,\eta(x_1))$ separates as before $\eta$ and $\zeta$. 

\smallskip

\noindent\textbf{Subcase: $x_o,x^o$ are two distinct points in common between the not separated $\eta$ and $\zeta$} (see Figure~\ref{fig:ConvexShape} bottom-left plot):\\
The line $y=mx+q$ connecting the two common points separates the two graphs $\eta$ and $\zeta$, and the proof is concluded.
\end{proof}

We now show a property of the sum of squares of the residual on a generic linear set of functions.

\begin{proposition}\label{prop:combin}
Let $(x_i,y_i)$, $i=1,\ldots,n$ be $n$ points and let $\Theta$ be a linear set of functions, which means that $a\theta_1+b\theta_2 \in \Theta$ whenever $\theta_1,\theta_2\in\Theta$.
If 
\[
\theta_0 = \arg\min_{\theta\in\Theta} \sum_{i=1}^n(y_i-\theta(x_i))^2,
\]
then for any $\theta_1\in\Theta$ and $t\in[-1,1]$,
\[
\sum_{i=1}^n(y_i-\theta_1(x_i))^2 \geq \sum_{i=1}^n\Big[y_i-\big(t\,\theta_1(x_i)+(1-t)\,\theta_0(x_i)\big)\Big]^2,
\]
and the inequality is strict if $\theta_1\neq \arg\min_{\theta\in\Theta} \sum_{i=1}^n(y_i-\theta(x_i))^2$ and $t\in(-1,1)$.

If $\Theta$ is a convex set, then the conclusion still holds with $t\in[0,1]$ and $t\in [0,1)$ in place of $t\in[-1,1]$ and $t\in (-1,1)$.
\end{proposition}

\begin{proof}[Proof of Proposition~\ref{prop:combin}.]
Given $\theta_0$ and $\theta_1$, note that 
\begin{multline*}
g(t) = \sum_{i=1}^n \!\Big[y_i-\big(t\,\theta_1(x_i)+(1-t)\,\theta_0(x_i)\big)\Big]^2 =
\sum_{i=1}^n\big[ t\,(\theta_0(x_i)-\theta_1(x_i)) + y_i-\theta_0(x_i)\big]^2 \!
\\
= t^2 \sum_{i=1}^n (\theta_0(x_i)-\theta_1(x_i))^2 +2t\sum_{i=1}^n (\theta_0(x_i)-\theta_1(x_i))(y_i-\theta_0(x_i)) + \sum_{i=1}^n (y_i-\theta_0(x_i))^2 
\end{multline*}
defines a parabola with minimum in $t=0$, since
$$
g(0) = \sum_{i=1}^n(y_i-\theta_0(x_i))^2 = \min_{\theta\in\Theta} \sum_{i=1}^n(y_i-\theta(x_i))^2.
$$
Then $g(1)=g(-1)$ and $g(t)< g(1)$ for any $t\in (-1,1)$ if $g(0)\neq g(1)$. When $\Theta$ is only a convex set, $t$ can be chosen in $[0,1]$ to guarantee that $t\theta_1+(1-t)\theta_0\in\Theta$.
\end{proof}

\begin{corollary}\label{cor:combin}
Let $(x_i,y_i)$, $i=1,\ldots,n$ be $n$ points and let $y=m_0x+q_0$ be the least square linear estimator. Then for any line $y=mx+q$ with $mm_0\leq 0$, there exists a constant $c$ such that
\[
\sum_{i=1}^n[y_i-(mx_i+q)]^2 \geq \sum_{i=1}^n(y_i-c)^2.
\]
The inequality is strict if $m\neq m_0$ or $q\neq q_0$.
\end{corollary}
\begin{proof}[Proof of Corollary~\ref{cor:combin}.]
Apply Proposition~\ref{prop:combin} to the set of lines, with $t^* = \vert m_0\vert /(\vert m\vert +\vert m_0\vert )$ ($0/0=0$). Since $m\vert m_0\vert + m_0\vert m\vert =0$, we get
$$
    \sum_{i=1}^n[y_i-(mx_i+q)]^2 \geq \sum_{i=1}^n\big[y_i-(t^*q+(1-t^*)q_0)\big]^2,
$$
which is the thesis with $c=t^*q+(1-t^*)q_0$.
\end{proof}

\begin{lemma}\label{lem:combin2}
% Let $(x_*,y_*)$ and $(x^*,y^*)$ be two points with $x_*<x^*$ and 
Let  $n_*$ and $n^*$ be two positive numbers, and assume $y_*\geq y^*$, $z_*<z^*$. Then
\begin{equation}\label{eq:nyz_*^*}
n_*(y_*-z_*)^2+n^*(y^*-z^*)^2 > n_*\Big(y_*-\frac{n_*z_*+n^*z^*}{n_*+n^*}\Big)^2 + n^*\Big(y^*-\frac{n_*z_*+n^*z^*}{n_*+n^*}\Big)^2,
\end{equation}
and in particular there exists a constant $c$ such that
\[
n_*(y_*-z_*)^2+n^*(y^*-z^*)^2 > n_*(y_*-c)^2 + n^*(y^*-c)^2. 
\]
\end{lemma}
\begin{proof}[Proof of Lemma~\ref{lem:combin2}.]
Call $\bar{z} =\frac{n_*z_*+n^*z^*}{n_*+n^*}$. We have
\begin{multline*}
    n_*(y_*-z_*)^2+n^*(y^*-z^*)^2 = n_*(y_*-\bar{z}+\bar{z}-z_*)^2+n^*(y^*-\bar{z}+\bar{z}-z^*)^2
    \\
     = n_*(y_*-\bar{z})^2+n^*(y^*-\bar{z})^2 
        +n_*(\bar{z}-z_*)(2y_*-\bar{z}-z_*)
        +n^*(\bar{z}-z^*)(2y^*-\bar{z}-z^*).
\end{multline*}
and then, since
\[
\bar{z}-z_* = \frac{n^*z^*-n^*z_*}{n_*+n^*} = n^*\frac{z^*-z_*}{n_*+n^*} , 
\qquad \bar{z}-z^* =\frac{n_*z_*-n_*z^*}{n_*+n^*} = n_*\frac{z_*-z^*}{n_*+n^*},
\]
we obtain
\begin{multline*}
    \Big[ n_*(y_*-z_*)^2+n^*(y^*-z^*)^2 \Big] - \Big[ n_*(y_*-\bar{z})^2+n^*(y^*-\bar{z})^2 \Big] \\
    \begin{aligned}
     & = n_*n^*\frac{z^*-z_*}{n_*+n^*}(2y_*-\bar{z}-z_*)
        +n_*n^*\frac{z_*-z^*}{n_*+n^*}(2y^*-\bar{z}-z^*)
    \\
    & = n_*n^*\frac{z^*-z_*}{n_*+n^*} \Big( 2y_*-\bar{z}-z_* - 2y^*+\bar{z}+z^*\Big).
    \\
    & = n_*n^*\frac{z^*-z_*}{n_*+n^*} \Big[ 2(y_*-y^*) + (z^*-z_*)\Big],
    \end{aligned}
\end{multline*}
which is positive, since $y_*\geq y^*$, $z_*<z^*$ and $n_*,n^*>0$.
\end{proof}

\section*{Acknowledgements}
This work was supported by the Italian national project PRIN 2022 [grant number 2022TRB44L]. Caterina May's work was supported also by UK Engineering and Physical Sciences Research Council grant
EP/T021624/1.
Giacomo Aletti is member of “Gruppo Nazionale per il Calcolo Scientifico (GNCS)” of the Italian Institute “Istituto Nazionale di Alta Matematica (INdAM)”.

% \bibliographystyle{Chicago}
% \bibliography{bias}% common bib file

\end{document}